\documentclass[aps,prl,twocolumn,english,balance,superscriptaddress,floats,showpacs,letterb]{revtex4-2}
\usepackage[latin9]{inputenc}
\usepackage{amsmath}
\usepackage{amssymb}
\usepackage{stmaryrd}
\usepackage{graphicx}
\usepackage{subfigure}
\usepackage{multirow}
\usepackage{mathtools}
\usepackage{xcolor}
\usepackage{gensymb}
\usepackage{calligra}

\DeclareMathAlphabet{\mathcalligra}{T1}{calligra}{m}{n}
\DeclareFontShape{T1}{calligra}{m}{n}{<->s*[2.2]callig15}{}

\newcommand{\eneq}{\end{equation}}
\newcommand{\beq}{\begin{equation}}
\newcommand{\eeq}{\end{equation}}
\newcommand{\bea}{\begin{eqnarray}}
\newcommand{\eea}{\end{eqnarray}}
\newcommand{\bwt}{\begin{widetext}}
\newcommand{\ewt}{\end{widetext}}

\newcommand{\eps}{\epsilon}
\newcommand{\bk}{\mathbf{k}}

\newcommand{\fvec}[1]{\boldsymbol{#1}}

\newcommand{\half}{\frac{1}{2}}

\newcommand{\rmd}{{\rm d}}

\makeatother

\begin{document}
	
\title{Cascades between light and heavy fermions in the normal state of magic angle twisted bilayer graphene}

\author{Jian Kang}
\email{jkang@suda.edu.cn}
\affiliation{School of Physical Science and Technology \& Institute for Advanced Study, Soochow University, Suzhou, 215006, China}

\author{B. Andrei Bernevig}
\affiliation{Department of Physics, Princeton University, Princeton, NJ 08544, USA}

\author{Oskar Vafek}
\email{vafek@magnet.fsu.edu}
\affiliation{National High Magnetic Field Laboratory, Tallahassee, Florida, 32310, USA}
\affiliation{Department of Physics, Florida State University, Tallahassee, Florida 32306, USA}

\begin{abstract}
We present a framework for understanding the recently observed cascade transitions and the Landau level degeneracies at every integer filling of twisted bilayer graphene. The Coulomb interaction projected onto narrow bands causes the charged excitations at an integer filling to disperse, forming new bands. If the excitation moves the filling away from the charge neutrality point, then it has a band minimum at the moire Brillouin zone center with a small mass that compares well with the experiment; if towards the charge neutrality point, then it has a much larger mass and a higher degeneracy. At a non-zero density away from an integer filling the excitations interact. The system on the small mass side has a large bandwidth and forms a Fermi liquid. On the large mass side the bandwidth is narrow, the compressibility is negative and the Fermi liquid is likely unstable. This explains the observed sawtooth features in compressibility, the Landau fans pointing away from charge neutrality as well as their degeneracies. By providing a description of the charge itineracy in the normal state this framework sets the stage for superconductivity at lower temperatures.
\end{abstract}
\maketitle

The discovery of the correlated insulating phases and superconductivity in the magic-angle twisted bilayer graphene has generated a flurry of research activity~\cite{BMModel,Pablo1,Pablo2,Cory1,David,Young,Cory2,Dmitry1,Yazdani,Ashoori,Dmitry2,Eva,Yazdani2,Shahal1,Young2,Stevan,YuanCao2020,Young3,Young20,Shahal2,StevanNature21,Young21,StevanNature21,Yacoby21,LiVafek,Eva21,PabloHunds,KangVafekPRX,LiangPRX1,Senthil1,Leon1,Kuroki,LiangPRX2,GuineaPNAS,BJYangPRX,Bernevig1,Leon2,Dai1,FengchengSC,Grisha,Stauber,KangVafekPRL,Bruno,Senthil2,SenthilC3,MacDonald,Zaletel1,Guinea2,Senthil3Ferro,Ashvin2,Sau,Zaletel2,Zaletel3,Chubukov,Dai2,YiZhang,KangVafekPRB,Ziyang,Fernandes2,Fengcheng,SenthilTop,LeonReview,Lucile,Zaletel4,VK2020,Cano,AndreiV,AndreiI,AndreiIII,AndreiIV,AndreiVI,ZaletelModes,ZaletelNematic,ZaletelSC,MacdonaldModes,TMI21, MacDonaldED}. This remarkable system exhibits correlated insulating phases at integer fillings of narrow bands~\cite{Pablo1,Pablo2,Cory1,David,Dmitry1,Young}, a hallmark of strong coupling physics.
Away from (certain) integer fillings, the same system becomes superconducting below a sufficiently low temperature, descending from a normal state exhibiting Fermi liquid-like quantum oscillations, both hallmarks of charge itineracy.

Recent observations of the cascade transitions in the compressibility and scanning tunneling microscopy studies at temperatures above the full onset of insulation or superconductivity~\cite{Shahal1,Yazdani2, Young20} have further sharpened this dichotomy. On the one hand, clear features associated with an integer filling of the moire unit cell were observed as expected in strong coupling~\cite{Cory2,Yazdani}. On the other hand, the electron system appears highly compressible when integer filling is approached from the charge neutrality point (CNP) side -- even with negative compressibility -- and much less compressible when approached from the remote bands side, producing sawtooth features in the inverse compressibility vs filling plots~\cite{Shahal1,Young20,Shahal2,Yacoby21,StevanNature21}. This led the authors of Ref.~\cite{Shahal1} to propose a simple ``Dirac revival'' picture based on the strictly intermediate coupling of a simplified model in which the non-interacting Bistritzer-MacDonald (BM)~\cite{BMModel} bands are sequentially filled. In this picture, starting from the CNP the BM bands are filled equally until a critical filling after which one of the flavors is nearly fully populated, while the densities of the remaining flavors are reset to somewhat below the CNP. The key source of itineracy for such a proposal is the dispersion of the BM bands. Unfortunately, the BM bands also feature {\em two} Dirac nodes per spin and valley, doubling the Landau level degeneracy away from each integer filling to $8,6,4,2$ sequence, and making this proposal inconsistent with the observed $4,3,2,1$ sequence.

Here we show that the non-trivial narrow band topology/geometry~\cite{Senthil1,SenthilTop,Bernevig1,BJYangPRX,Dai1}, neglected in the simplified model of Ref.~\cite{Shahal1}, combined with Coulomb interaction can drive the itineracy of the single particle charge excitations near the integer fillings even in strong coupling, i.e.~when the BM kinetic energy is neglected. In addition to insulating phases belonging to spin-valley U(4) or U(4)$\times$U(4) manifold~\cite{KangVafekPRL,Zaletel3,AndreiIV}, the interplay of band topology/geometry and strong Coulomb interactions was shown to make the strong coupling nematic phases, which are semi-metallic, energetically competitive~\cite{Ashvin2,KangVafekPRB}. The nematic phase was recently shown to be further stabilized by strain~\cite{ZaletelNematic}. Absence of gaps is therefore not at variance with the strong coupling picture.

\begin{widetext}
\begin{figure*}[t]
	\includegraphics[scale=0.28]{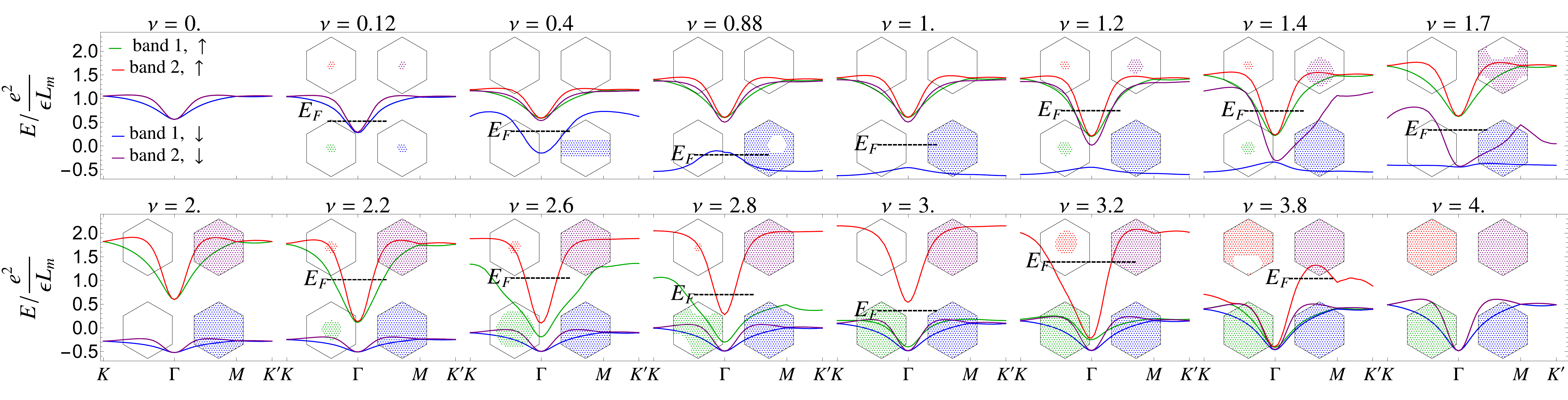}
	\includegraphics[scale=0.28]{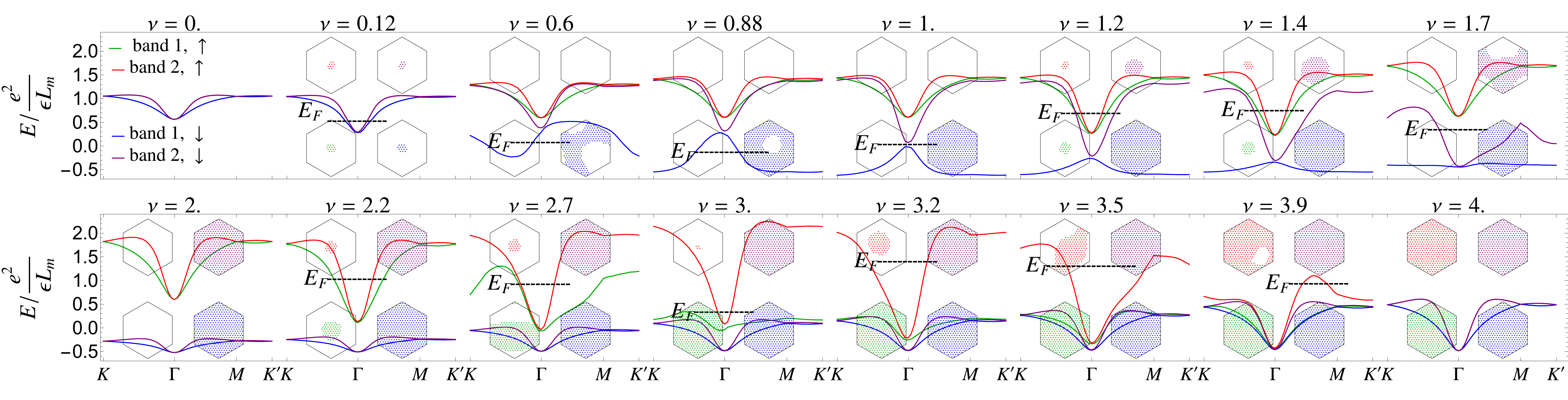}
\caption{Evolution of the quasiparticle bands upon the change of the filling factor $\nu$ for the trial state in Eqn.~\ref{Eqn:trial} at $w_0/w_1=0.7$ when the $C_2 \mathcal{T}$ symmetry is allowed to be broken (top two panels) and when $C_2\mathcal{T}$ is enforced (bottom two panels). The hexagonal insets show the occupied $\fvec k$ states for each of the bands.}
	\label{Fig:BandEvolution}
\end{figure*}
\end{widetext}

\begin{figure}[t]
	\includegraphics[scale=0.28]{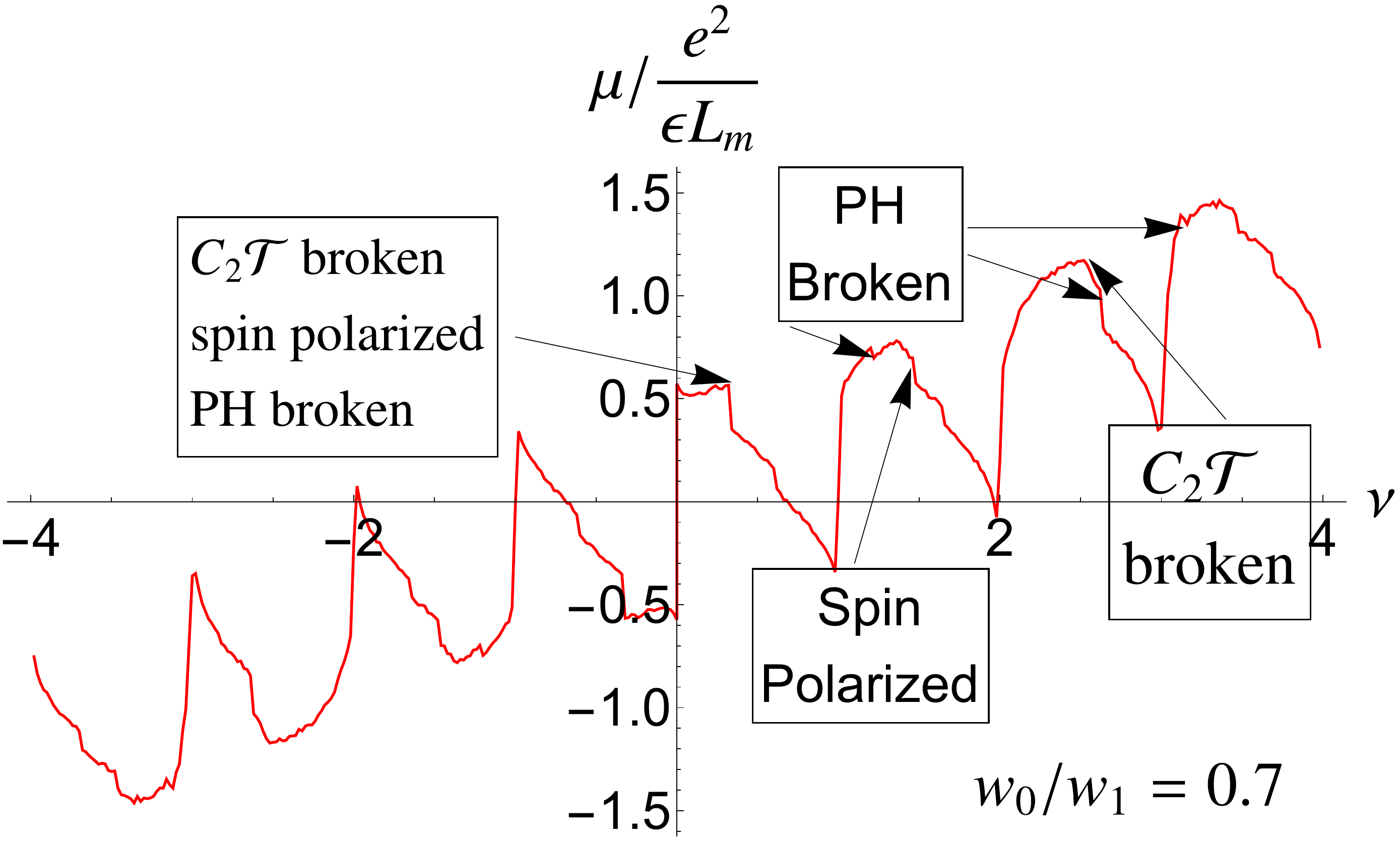}
	\includegraphics[scale=0.28]{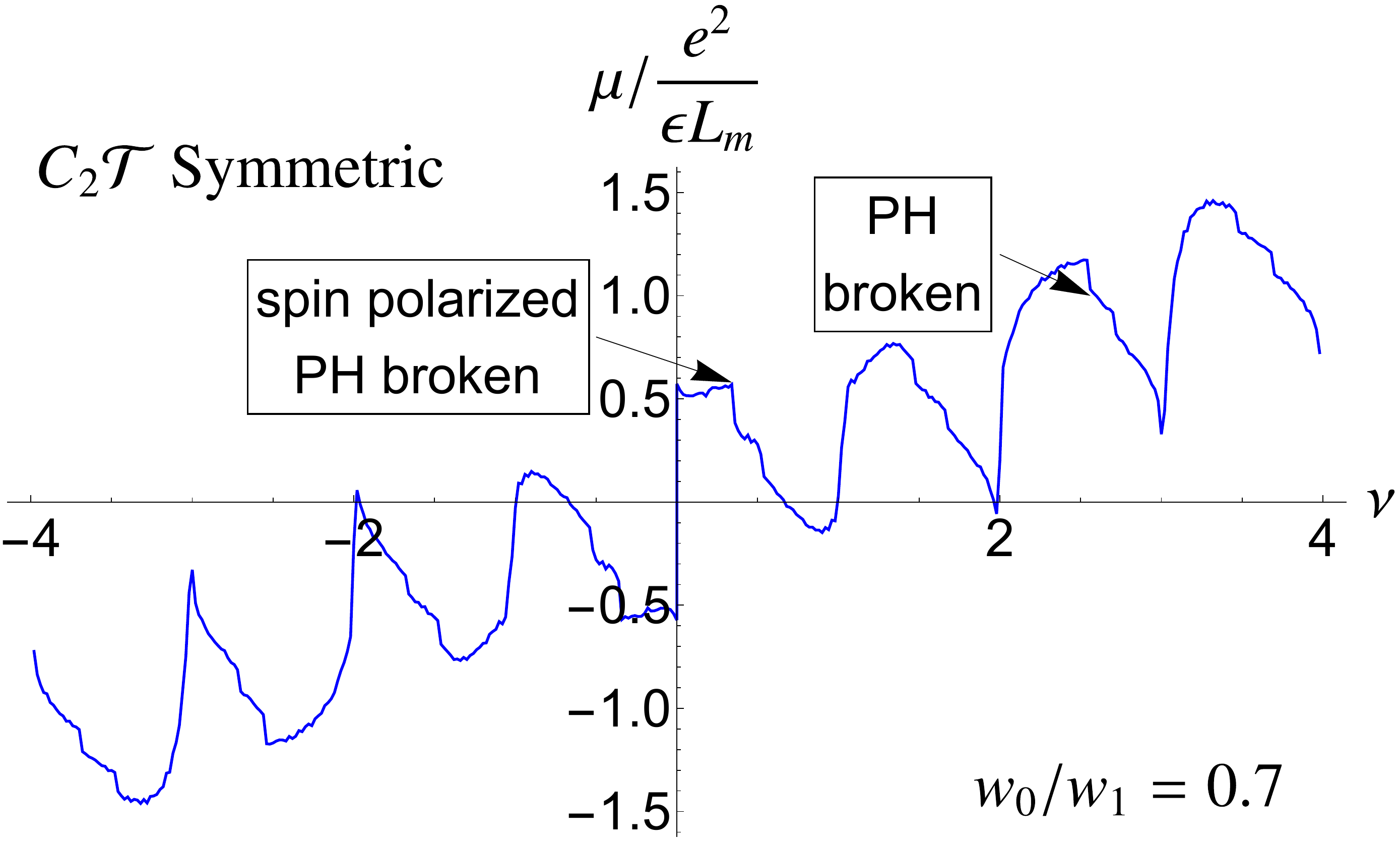}
\caption{Evolution of the chemical potential $\mu$ as the filling $\nu$ varies between $-4$ and $4$ when $C_2\mathcal{T}$ symmetry is allowed to be broken (top panel) and when $C_2\mathcal{T}$ symmetry is enforced (bottom panel).}
	\label{Fig:MuNu}
\end{figure}

Interestingly, in all of these phases, whether insulating or semi-metallic, the band minimum of the single particle charge excitations appears at $\fvec \Gamma$, the center of the moire Brillouin zone (mBZ), naturally producing the experimentally observed sequence of weak magnetic field Landau level degeneracies. Here we provide an explanation of this observation and find that the strong coupling band degeneracies are a consequence of a novel action of the combination of the unitary particle-hole~\cite{Bernevig1} and the $C_2\mathcal{T}$ symmetries. We find that the band dispersion of a single particle or a single hole added to the strong coupling phases at a non-zero integer filling is highly asymmetric (see Fig.~\ref{Fig:BandEvolution}). If the excitation moves the filling closer to (away from) the CNP it is heavy with a narrow bandwidth (light with a large bandwidth). The light mass excitations have a minimum at $\fvec \Gamma$ and a smaller degeneracy than the heavy ones, whose minima are away from a high symmetry $\fvec k$-point. At a finite density away from an integer filling, the single particle excitations repel each other~\cite{AndreiV}. By estimating the ratio of the residual interaction to the kinetic energy obtained by filling the new (non-rigid) bands, the system on the small mass side is effectively in the Fermi liquid phase.
We find that the value of the mass here compares favorably with the existing experiments without adjustable parameters\cite{SM}.
On the heavy mass side, we found several nearly degenerate states that are related by many particle-hole 
excitations, suggesting that there, the residual interactions may lead to additional instabilities of a heavy Fermi liquid. This explains the observed Landau fans pointing away from the CNP as well as their degeneracies. The computed chemical potential $\mu$ displays features similar to experimental observations. This includes negative compressibilities and the overall magnitude of its difference between fully occupied and empty eight narrow bands, regardless of whether the strong coupling states at odd integer filling are gapped or gapless (see Fig.~\ref{Fig:MuNu}).

Our starting Hamiltonian includes only the momentum conserving Coulomb interactions (renormalized by the remote bands) projected onto the BM narrow bands
\begin{align}
H = \frac1{2 A} \sum_{\fvec q \neq 0} V(\fvec q) \delta \rho_{\fvec q} \delta \rho_{-\fvec q}. \label{Eqn:HInt}
\end{align}
Here $A$ is the area of the system, $V(\fvec q) =(\epsilon q/(2 \pi e^2) + \Pi(q))^{-1}$, for the encapsulating hexagonal boron-nitrite $\epsilon = 4.4$, and the static polarization function $\Pi(q)$ originates from the remote bands~\cite{SM}. $\delta \rho_{\fvec q} = \rho_{\fvec q} - \bar{\rho}_{\fvec q}$ is the difference between the projected density operator and the background charge density, and  $\fvec q$ is not restricted to the first mBZ (unlike the sum over $\fvec k$ below). Specifically,
\begin{align}
 \rho_{\fvec q} & = \sum_{\substack{\tau = \fvec K, \fvec K' \\ s = \uparrow \downarrow}} \sum_{\substack{\fvec k \in \mathrm{mBZ} \\ n, n' = \pm}} \Lambda_{n n'}^{\tau}(\fvec k, \fvec k+ \fvec q) d^{\dagger}_{\tau, n, s, \fvec k} d_{\tau, n', s, \fvec k + \fvec q} \label{Eqn:Rhoq} \\
 \bar{\rho}_{\fvec q} & =  2 \sum_{\fvec G, n = \pm} \delta_{\fvec q, \fvec G} \sum_{\fvec k \in \mathrm{mBZ}} \Lambda^{\fvec K}_{nn}(\fvec k, \fvec k + \fvec G) \label{Eqn:RhoBar} \ ,
\end{align}
where the projected density operator $\rho_{\fvec q}$ is expressed in the Chern basis $\Phi_{\tau, \pm, \fvec k}(\fvec r)$ that carries the indices of the valley $\tau = \fvec K$ or $\fvec K'$, the Chern $n = \pm$, the spin $s = \uparrow \downarrow$, and the  $\fvec k$, with the corresponding creation and annihilation operators $d^{\dagger}$ and $d$. The Chern states are the sublattice polarized states of the BM model for narrow bands~\cite{Zaletel3,KangVafekPRB} at the magic angle i.e. $w_1/(v_F k_{\theta}) = 0.586$ and $w_0/w_1 = 0.7$, where $w_0$ and $w_1$ are the two interlayer couplings~\cite{LiangPRX1,Senthil1,Grisha}, $v_F$ is the Fermi velocity for the monolayer graphene, $k_{\theta} = 8\pi/(3 L_m) \sin(\theta/2)$ and $L_m$ is the moire lattice constant. Spinless time reversal symmetry relates the valleys $\fvec K$ and $\fvec K'$~\cite{Senthil1,KangVafekPRX,LiangPRX1}.
The form factor matrix 
\begin{align*}
\Lambda^{\tau}_{mn}(\fvec k, \fvec k + \fvec q) = \int_{uc}\rmd \fvec r\ e^{-i \fvec q \cdot \fvec r} \Phi^*_{\tau, m, \fvec k}(\fvec r) \Phi_{\tau, n, \fvec k + \fvec q}(\fvec r) 
\end{align*}
contains the information about the non-trivial topology/geometry of the narrow bands and plays an important role in the physics we describe; it has been neglected in Ref.~\cite{Shahal1}. 

Previous analytical and numerical works have established that over a large range of parameters the ground states $| \Psi_{GS} \rangle$ of the Hamiltonian $H$ in Eqn.~\ref{Eqn:HInt} are Slater determinants~\cite{KangVafekPRL,Zaletel3,KangVafekPRB,AndreiVI,AndreiIV}. At even integer fillings they consist of all states that satisfy~\cite{Zaletel3,AndreiIV,VK2020}
\begin{equation}
\delta \rho_{\fvec q} | \Psi_{GS} \rangle = \frac{\nu}4 \sum_{\fvec G} \delta_{\fvec q, \fvec G} \bar{\rho}_{\fvec G} | \Psi_{GS} \rangle \  ,   \label{Eqn:DeltaRho}
\end{equation}
with the eigenenergy $E_{\nu} = \frac1{2A} \sum_{\fvec G \neq 0} V(\fvec G) \left| \frac{\nu}4 \bar{\rho}_{\fvec G} \right|^2$.
The exact excited states can also be obtained~\cite{VK2020,AndreiV}. Indeed, acting with $H$ on the state $\hat{X}| \Psi_{GS} \rangle$, where $\hat{X}$ is some combination of $d^\dagger$s and $d$s, and using (\ref{Eqn:DeltaRho}), we find an eigenequation
\begin{align}
& (H - E_{\nu}) \hat{X} | \Psi_{GS} \rangle = \frac1{2A} \sum_{\fvec q} V(\fvec q) \left( [\delta \rho_{-\fvec q}, [\delta \rho_{\fvec q}, \hat{X}]] + \right. \nonumber \\
& \quad \left. [\delta \rho_{\fvec q}, \hat{X} ] \delta \rho_{-\fvec q} + [\delta \rho_{-\fvec q}, \hat{X} ] \delta \rho_{\fvec q} \right)| \Psi_{GS} \rangle \ . \label{Eqn:HX}
\end{align}
The last two terms can be further simplified by applying Eqn.~\ref{Eqn:DeltaRho}. Because each commutator has the same number of $d^{\dagger}$s and $d$s as the ones in $\hat{X}$, we can readily match the coefficients. This was used to find the charge neutral collective modes\cite{VK2020,AndreiV} and to show that the spectrum of charge-2 elementary excitations for a purely repulsive $V(\fvec q)$ does not have a bound state~\cite{AndreiV}.
For $\hat{X}_+=d^\dagger_{\tau,n,s,\fvec{k}}$ and $\hat{X}_-=d_{\tau,n,s,\fvec{k}}$,  Eqn.(\ref{Eqn:HX}) reduces to solving for eigenvalues of the  2$\times$2 matrix
\begin{eqnarray}
&&\mathcal{E}^{\tau}_{n' n,\pm}(\fvec k) = \frac1{2 A}   \left( \sum_{\fvec q} V(q) \sum_m \Lambda^{\tau}_{mn}(\fvec k - \fvec q, \fvec k) \Lambda^{\tau}_{n' m}(\fvec k, \fvec k - \fvec q)  \right. \nonumber \\
&& \quad \left.  \pm \frac{\nu}2 \sum_{\fvec G} V(G) \bar{\rho}_{\fvec G} \Lambda^{\tau}_{n'n}(\fvec k + \fvec G, \fvec k ) \right) , \label{Eqn:ExEne}
\end{eqnarray}
that leads to $2$ different bands for both electron and hole excitations for each spin $s$.
To illustrate the main effect, let us first consider the chiral limit~\cite{Grisha,Niu2020,Becker2020}, $w_0/w_1 = 0$. In this case the Chern states are perfectly sublattice polarized. Therefore, $\Lambda^{\tau}_{mn}(\fvec k, \fvec k + \fvec q)$ is diagonal in $m,n$ and Slater determinant states obtained by filling Chern bands satisfy (\ref{Eqn:DeltaRho}) also at odd filling; they have been shown to be the ground states in exact diagonalization (ED) studies in Ref.~\cite{AndreiVI}. Consequently, the spectrum of the single particle excitations can be solved with the Eqn.~\ref{Eqn:ExEne} at any integer filling. Interestingly, the eigenstates of $\mathcal{E}^{\tau}_{n' n,+}(\fvec k)$ are exactly degenerate over the whole mBZ, as are the eigenstates of $\mathcal{E}^{\tau}_{n' n,-}(\fvec k)$. This is due to the combination of the 2-fold rotation about the axis normal to the plane, spinless time reversal and the chiral particle-hole symmetries~\cite{Grisha,AndreiIII,Zaletel3,Cano}, $\mathcal{K}'=C_2\mathcal{T}\mathcal{C}$. Because $\mathcal{K}'$ preserves $\bk$ and $\mathcal{K}'^2=-1$, $\mathcal{E}^{\tau}_{n' n,\pm}(\fvec k)$ must be proportional to $\delta_{mn}$ for each $\fvec k$.

Moving away from the chiral limit, i.e. $w_0/w_1\neq 0$, we see that the particle and hole dispersions are the same at the CNP, as can be understood from Eqn.~\ref{Eqn:ExEne} for $\nu = 0$. In addition, the two bands are now degenerate only at high symmetry points $\fvec \Gamma$, $\fvec M$, $\fvec K$ and $\fvec K'$ (see Fig.~\ref{Fig:BandEvolution}). The degeneracies at $\fvec \Gamma$ and $\fvec M$ are protected by the combination of $C_2\mathcal{T}$ and particle-hole symmetry $\mathcal{P}$ that is discussed in Ref.~\cite{Bernevig1,VK2020,SM}.  Moreover, combined with $C_3$ symmetry, the winding numbers at $\fvec \Gamma$ and $\fvec M$ can be shown to be $3$ and $-1$ respectively. The degeneracy at $\fvec K$ (and $\fvec K'$) is protected by $C_3$ with the winding number of $1$ (see Ref.~\cite{SM}). 

Although such degeneracy and winding numbers are also seen at other even integer fillings, $\nu=\pm 2,\pm 4$, excitation spectra are markedly different. The bands away from CNP have the minimum at $\fvec \Gamma$ and the bandwidth of the order of the Coulomb scale $e^2/(\epsilon L_m)$. However, the bands towards CNP are rather flat and have their minima away from high symmetry $\fvec k$ points. 
To understand the origin of this effect, we return to the chiral limit ($w_0/w_1=0$) and analyze the first (exchange) and the second (direct) terms in the Eqn.~\ref{Eqn:ExEne}.
Both of these terms can be well approximated by a nearest neighbor (NN) tight-binding model on a triangular lattice with a negative NN hopping amplitudes $t_E=-0.0551\frac{e^2}{\eps L_m}$ and $t_D=-0.0544\frac{e^2}{\eps L_m}$,  and with onsite terms $\eps_E=1.731\frac{e^2}{\eps L_m}$ and $\eps_D=0.326\frac{e^2}{\eps L_m}$ for exchange and direct terms respectively~\cite{VK2021}. This, as well as our $\fvec{k}\cdot\fvec{p}$ analysis based on the model in Ref.~\cite{AndreiI,SM}, show that the minimum of the dispersion is at $\fvec \Gamma$ when the two terms add. When they subtract, the bandwidth is reduced. 
Note that the magnitudes of the NN hoppings $t_E$ and $t_D$ are such that at $\nu=\pm 1$ the cancellation is nearly complete, leading to the narrow band of heavy holes at $\nu=1$ and heavy particles at $\nu=-1$.
Accordingly, for $|\nu| \geq 2$, the dispersions towards CNP reverse compared to $\nu=0$, also with heavy excitations.
Because for excitations at $\nu\neq 0$ that are moving the filling away from the CNP the direct and the exchange terms add (in absolute value), the resulting bands are more dispersive with a minimum at $\fvec \Gamma$. These are the light fermions.
As seen in Fig.\ref{Fig:BandEvolution}, the effect persists away from the chiral limit $w_0/w_1\neq 0$.

At a finite density away from an integer filling the excitations interact\cite{VK2020,AndreiV} with each other as can be seen from Eq.~\ref{Eqn:HX}. Nevertheless, the steep dispersion observed for a {\it single} electron (hole) added to the exact eigenstates at the positive (negative) integer fillings and at CNP suggests that at a finite density close to the integer filling -- and in the direction away from CNP -- the kinetic energy of such excitations is sufficient to stabilize a Fermi liquid. This is broadly consistent with the ED results of Ref.~\cite{Bergholtz2020}, where emergent Fermi liquids  were also found in different, albeit related, models of moire heterostructures.
We therefore approximate the ground state by the trial state 
\begin{equation}
    |\Psi_{GS} \rangle = \prod_{s, \fvec k} \hat{P}_{s, \fvec k} | \Psi_{CNP} \rangle , \label{Eqn:trial}
\end{equation}
where $| \Psi_{CNP} \rangle$ is a ground state at CNP which, without loss of generality, is taken to be completely $\fvec K'$ valley polarized with all four $\fvec K$ bands empty. At each $s$, $\fvec k$ there are two bands at $\fvec K$ whose occupation number is denoted by $\nu_{s,\fvec k}$; when empty ($\nu_{s, \fvec k} = 0$)  $\hat P_{s, \fvec k}=1$ and when doubly occupied ($\nu_{s, \fvec k} = 2$) $\hat{P}_{s, \fvec k}  = d^{\dagger}_{\fvec K, +, s, \fvec k} d^{\dagger}_{\fvec K, -,s, \fvec k}$. 
When singly occupied ($\nu_{s, \fvec k} = 1$), we have $\hat{P}_{s, \fvec k} = u_{s, \fvec k} d^{\dagger}_{\fvec K, + , s, \fvec k} + v_{s, \fvec k} d^{\dagger}_{\fvec K, - , s, \fvec k}$  with variational parameters satisfying 
$|u_{s, \fvec k}|^2 + |v_{s, \fvec k}|^2 = 1$.
The integer parameters $\nu_{s, \fvec k}$ are also determined variationally and satisfy the constraint $\sum_{s, k} \nu_{s, \fvec k}=\nu N_{uc}$, where $N_{uc}$ is the total number of moire unit cells. Minimizing $E = \langle \Psi_{GS} | H | \Psi_{GS} \rangle$ subject to the mentioned constraints yields the self-consistent eigen-equations for $u_{s,\fvec k}$ and $v_{s,\fvec k}$
\begin{equation}
     H_{eff}(s, \fvec k) \begin{pmatrix}  u_{s, \fvec k} \\ v_{s, \fvec k}  \end{pmatrix}  = E_{\alpha}(s, \fvec k) \begin{pmatrix}  u_{s, \fvec k} \\ v_{s, \fvec k}  \end{pmatrix} \label{Eqn:EffBands}.
\end{equation}
The effective Hamiltonian $H_{eff}$ is discussed in detail in SM~\cite{SM}. $E_{\alpha}(s, \fvec k)$ specifies the band structure shown in Fig.~\ref{Fig:BandEvolution}. Fig.~\ref{Fig:MuNu} shows the filling dependence of chemical potential $\mu$, calculated from the constraint $\sum_{\alpha, s, \fvec k} \Theta(\mu - E_{\alpha}(s, \fvec k))  = \nu N_{uc}$. The following discussion focuses on $\nu \geq 0$,  the states with $\nu < 0$ can be obtained using the many-body particle-hole symmetry~\cite{AndreiIII}.

At $\nu = 2$, our variational method results in  $| \Psi_{GS}^{\nu = 2} \rangle = \prod_{n = \pm, \fvec k} d^{\dagger}_{\fvec K, n, s, \fvec k} | \Psi_{CNP} \rangle$ where the spin $s = \uparrow$ or $\downarrow$. Although this exact (gapped) eigen-state breaks the time reversal symmetry (spinfull and spinless), it does not break $C_2\mathcal{T}$. Thus it carries zero Chern number. It was also numerically shown to be the ground state~\cite{AndreiVI}. Its single particle excitation spectrum produced by Eqn.~\ref{Eqn:EffBands} is the same as the ones obtained in Eqn.~\ref{Eqn:ExEne}. 
At odd integer fillings with $w_0/w_1 = 0.7$ this method results in the quantum anomalous Hall (QAH) state with spontaneously broken $C_2\mathcal{T}$ symmetry if no additional constraints are applied as shown in the upper two panels of Fig.~\ref{Fig:BandEvolution}. This result is consistent with the exact solution obtained in the chiral limit ($w_0/w_1 = 0$), the recent DMRG calculation~\cite{KangVafekPRB,Zaletel4} and the ED~\cite{AndreiVI} for a range of $w_0/w_1\neq 0$. For comparison, applying the $C_2\mathcal{T}$ symmetric constraint to the odd filling trial state $|\Psi_{GS}\rangle$ leads to a semi-metallic nematic state as shown in the lower two panels of Fig.~\ref{Fig:BandEvolution}. Both the $C_2\mathcal{T}$ broken Chern insulators and $C_2\mathcal{T}$ symmetric gapless states are nearly degenerate, as also demonstrated by DMRG and ED calculations~\cite{KangVafekPRB,Zaletel4,AndreiVI}. 

At non-integer fillings $|\Psi_{GS}\rangle$ leads to gapless compressible phases.
The details of the band evolution with filling are shown in Fig.~\ref{Fig:BandEvolution}.
At fillings just above the positive integers the gapless excitation spectrum can be seen to be strongly dispersive, with the bandwidth of the order of the Coulomb interaction. As discussed below, we expect such low compressibility phases to be stable when the residual interaction that scatters among different trial states is included, resulting in Fermi liquids at these fillings. 
The ultimate instability of the Fermi liquids upon approaching a positive integer filling from below stems from the mentioned residual interactions {\it and} the fact that the band structure is not rigid, with the partially filled band(s) flattening as $\nu$ approaches an integer (see Fig.~\ref{Fig:BandEvolution}).
Even within this simple variational method, which does not account for the residual interactions, there are several Stoner-like phase transitions as the integer filling is approached from below. 
Such spontaneous breaking of $C_2\mathcal{T}$, particle-hole, or $C_3$ symmetries, furthers the instabilities of the Fermi liquid. We found the transition occurring between $\nu=0$ and $\nu=1$ to be first order, becoming a second order between higher integers fillings. 

As illustrated in Fig.~\ref{Fig:MuNu}, at each non-negative integer $\nu$, the chemical potential $\mu$ increases as $\nu$ moves away from the CNP. Before $\nu$ gets to the next integer, $\mu$ reaches its local maximum at a fractional filling and then decreases, resulting in the negative compressibilty $\frac{\rmd \mu}{\rmd\nu}$. The net increase of $\mu$ is $\sim 40$meV which compares well with $\sim 50$meV found in experiments~\cite{Ashoori,Yazdani2,Shahal1,Young20,Yacoby21}.

Because the dominant residual interaction is repulsive~\cite{VK2020,AndreiV}, we estimate its importance over dispersion in two different ways. {\it First}, we consider $r_s$, defined as the ratio of $U(\bar{r}) = \int \frac{\rmd^2 \fvec q}{(2\pi)^2} V(q) e^{i \fvec q \cdot \bar{\fvec r}}$, i.e. the residual Coulomb potential energy of two excitations separated by ${\bar r}=1/\sqrt{\delta n}$, and the average kinetic energy $E_K$; here $\delta n$ is the density deviation from the closest integer filling. For an electron excitation of a partially filled band we define $E_K^e = \int_{filled} \frac{\rmd^2 \fvec k}{(2\pi)^2} \left( E(\fvec k) - E_{\min} \right)$ where $E_{\min}$ is the band minimum, while for hole excitations, $E_K^h = \int_{unfilled} \frac{\rmd^2 \fvec k}{(2\pi)^2} \left( E_{\max} - E(\fvec k) \right)$ where $E_{\max}$ is the band  maximum. Then, $E_K$ is set to be the smaller of $E_K^e$ and $E_K^h$.  As $\nu$ approaches an integer, $\delta n \rightarrow 0$ and $r_s = U(\bar{r})/ E_K$ diverges because $U(\bar{r}) \sim O(\sqrt{\delta n})$ and $E_K \sim O(\delta n)$. 
For $m<\nu \lesssim m+0.017$ where $m$ is a non-negative integer, we find  $r_s\geq 35$, i.e. $r_s$ is above the critical value for the Wigner crystallization~\cite{MC2D,Kivelson}. If we include additional screening due to the nearby metallic gates, $U(r)$ is modified from $1/r$ at long distances and decays faster when $r$ is larger than the distance to gates $l_g$. Therefore $U({\bar r})\ll E_K$ at small $\delta n$, eliminating a possible Wigner crystal if $\delta n < l_g^{-2}$. For a typical gate distance $l_g \sim 40$nm, the screened Coulomb interaction eliminates the Wigner crystal if $m<\nu\lesssim m+0.09$. Therefore, no Wigner crystal should exist close to an integer filling on the side away from the CNP.

{\it Second}, we calculate the ratio between $U({\bar r})$ and $W$, the bandwidth of the excitations. If $m <  \nu \lesssim m + 0.3$, then $U({\bar r})/W \lesssim 0.3$, suggesting that the system is in the weak coupling regime. Together with the above analysis of $r_s$, we conclude that the system is in the Fermi liquid phase if the filling is in this interval.  Moreover, as illustrated in Fig.~\ref{Fig:BandEvolution}, in this filling interval the $4 - m$ partially occupied bands are filled equally near $\fvec \Gamma$, resulting in the experimentally observed Landau fan degeneracy of $4 - m$ when pointing away from the CNP~\cite{Pablo1,Pablo2,Cory1,Dmitry1}. 

On the other hand, for $m+0.4 \lesssim  \nu < m + 1$, the variational calculation resulted in the band reconstruction and several nearly degenerate states. These states are related by many particle-hole excitations, implying that the obtained ground state, as well as the associated excitation spectrum, may be unstable upon including the residual interactions between the quasi-particles. Moreover, as discussed, the bands are narrow at every integer filling for excitations towards the CNP. This naturally explains the absence of the Landau fans towards the CNP~\cite{Pablo1,Pablo2,Cory1,Dmitry1}.

The framework presented here provides a strong coupling description of the itinerant carriers, whose residual interactions {\it and} dispersion both depend on the Coulomb interaction. The description of the charge itineracy presented here is in quantitative agreement with experiments, and builds a framework within which superconductivity, emerging at lower temperatures at some fillings, should be understood.
\begin{acknowledgments}
J.~K.~acknowledges the support from the NSFC Grant No.~12074276, and the Priority Academic Program Development (PAPD) of Jiangsu Higher Education Institutions. B.~A.~B.~is supported by the ONR No. N00014-20-1-2303 and partially by DOE Grant No. DE-SC0016239, NSF- MRSEC No. DMR-1420541 and DMR-2011750. O.~V.~is supported by NSF DMR-1916958 and partially by the National High Magnetic Field Laboratory through NSF Grant No.~DMR-1157490 and the State of Florida. 
This research was facilitated by the KITP program ``Correlated Systems with Multicomponent Local Hilbert Spaces'', supported in part by the National Science Foundation under Grant No. NSF PHY-1748958.

\end{acknowledgments}

\appendix

\begin{widetext}
	
	\newpage	
	
	\begin{center}
		\textbf{\large Supplemental Material for ``Cascades between light and heavy fermions in magic angle twisted bilayer graphene''}
	\end{center}
	
	\setcounter{equation}{0}
	\setcounter{figure}{0}
	\setcounter{table}{0}
	\makeatletter
	\renewcommand{\thefigure}{S\arabic{figure}}

\section{Renormalized Coulomb interaction}
In this section, we derive the formula of the renormalized Coulomb interaction. As discussed in the main text, the interaction can be written as $V(q)^{-1} = \epsilon q/(2\pi e^2) + \Pi(q)$, where $\Pi(q)$ is the static polarization function originated from the states at the remote bands. For states with the energy of $|E|$ much larger than the interlayer coupling $w_0$ and $w_1$, the dispersion can be approximated as that of a Dirac cone, and thus the Green function is
\[ G(i \omega, \fvec k) \approx \frac{\frac12 (1 + \sigma \cdot \hat{\fvec k})}{-i \omega + v_F k}  + \frac{\frac12 (1 - \sigma \cdot \hat{\fvec k})}{-i \omega - v_F k} \ . \]
By the random phase approximation, the polarization function  $\Pi_0(\fvec q)$ for these states on a single Dirac cone can be written as
\begin{align}
    \Pi_0(\fvec q) = - \int\frac{\rmd^2 k}{(2\pi)^2} \int\frac{\rmd \omega}{2\pi}  \mathrm{Tr}\left( G(i \omega, \fvec k_-) G(i \omega, \fvec k_+)  \right) \Theta(v_F k_- - E_c^*)  \Theta(v_F k_+ - E_c^*)
\end{align}
where $\fvec k_{\pm} = \fvec k \pm \half \fvec q$, $E_c^* = 0.15 v_F k_{\theta}$ is the low energy cutoff for the states on the remote bands, and $\Theta(v_F k_{\pm} - E_c^*)$ is the step function. Note that the terms  containing $\omega$ are
\[   \int \frac{\rmd \omega}{2 \pi} \frac1{(- i \omega \pm v_F k_-)(- i \omega \pm v_F k_+)} \]
The  integral vanishes if the two poles $\pm i v_F k_-$ and $\pm i v_F k_+$ are on the same side of the real axis. Thus, we found 
\begin{align}
    \Pi_0(\fvec q) = & \frac1{4v_F} \int \frac{\rmd^2 k}{(2\pi)^2} \frac{\Theta(v_F k_- - E_c^*)  \Theta(v_F k_+ - E_c^*)}{k_+ + k_-} \mathrm{Tr} \left(  (1 + \fvec \sigma \cdot  \hat{\fvec k}_+) (1 - \fvec \sigma \cdot  \hat{\fvec k}_-) + (1 + \fvec \sigma \cdot  \hat{\fvec k}_-) (1 - \fvec \sigma \cdot  \hat{\fvec k}_+) \right) \nonumber \\
    = & \frac1{v_F} \int \frac{\rmd^2 k}{(2\pi)^2} \frac{\Theta(v_F k_- - E_c^*)  \Theta(v_F k_+ - E_c^*)}{k_+ + k_-} \left( 1 - \hat{\fvec k}_+ \cdot \hat{\fvec k}_- \right)
\end{align}
With $\fvec q = 0$, $\fvec k_+ = \fvec k_- = \fvec k$, and thus $1 - \hat{\fvec k}_+ \cdot \hat{\fvec k}_- = 0$. Therefore, the polarization function $\Pi_0(0)$ also vanishes.

To further simplify the integral with non-zero $\fvec q$, we introduce two variables $y \in [1, \infty)$ and $\psi \in [0, 2\pi)$ so that
\[   k_{\parallel} = \frac{q}2 y \cos\psi \ , \quad k_{\perp} = \frac{q}2 \sqrt{y^2 -1} \sin \psi \ , \quad  \int \rmd^2 \fvec k = \int \rmd k_{\parallel} \ \rmd k_{\perp} = \int_1^{\infty} \rmd y \int_0^{2\pi} \rmd \psi \frac{q^2}4 \frac{y^2 - \cos^2 \psi}{\sqrt{y^2 -1}} \ ,
\]
where $k_{\parallel}$ and $k_{\perp}$ are the components of $\fvec k$ parallel and perpendicular to $\fvec q$. After the change of integral variables, we find
\[  \Pi_0(\fvec q) = \frac{2 q}{v_F} \int_0^{\frac{\pi}2} \frac{\rmd \psi}{2\pi} \sin^2\psi \int_{\max(1, \frac{2E_c^*}{v_F q} + \cos\psi)}^{\infty} \frac{\rmd y}{2\pi} \frac1{y\sqrt{y^2 - 1}}      \]
Then, we introduce $y = \sec\theta$ to obtain the analytic expression of the integral over $y$. For notation convenience, define $z = 2 E_c^*/(v_F q)$. The static polarization function $\Pi_0(\fvec q)$ is found to be
\[ \Pi_0(\fvec q) = \left\{ \begin{array}{ll}
     \dfrac{2q}{\pi v_F} \displaystyle\int_0^{\cos^{-1}(1 - z)} \dfrac{\rmd \psi}{2 \pi} \sin^2\psi \left( \tan^{-1} \sqrt{\dfrac{z + \cos\psi + 1}{z + \cos\psi - 1}} - \dfrac{\pi}4 \right) + \frac{q}{2 v_F} \int_{\cos^{-1}(1 - z)}^{\frac{\pi}2}  \frac{\rmd \psi}{2 \pi} \sin^2\psi  &   \mathrm{If}\ 0 < z < 1 \\
     \dfrac{2q}{v_F} \displaystyle\int_0^{\frac{\pi}2} \dfrac{\rmd \psi}{2 \pi} \sin^2\psi \left( \tan^{-1} \sqrt{\frac{z + \cos\psi + 1}{z + \cos\psi - 1}} - \dfrac{\pi}4 \right)  &   \mathrm{If}\  z > 1
\end{array}   \right.   \ . \]
In the above calculation, we considered the contribution of only one Dirac cone. Due to the spin, valley, and layer degree of freedom in the twisted bilayer graphene, the states of the remote bands are located on $8$ Dirac cones with the same $v_F$. Therefore, the polarization function originated from the states on the remote bands are $\Pi(\fvec q) = 8\Pi_0(\fvec q)$.

\section{Variational Method}
In this section, we discuss the variational method to obtain the ground states at the generic fillings. As discussed in the main text, the ground state is approximated as $|\Psi_{GS} \rangle = \prod_{s, \fvec k} \hat P_{s, \fvec k} | \Psi_{CNP} \rangle$, where $| \Psi_{CNP} \rangle$ is the ground state at $\nu = 0$. At each $s$, $\fvec k$, the operator $\hat P_{s, \fvec k}$ is $1$ if the occupation number $\nu_{s, \fvec k} = 0$, and $d^{\dagger}_{\fvec K, + , s, \fvec k} d^{\dagger}_{\fvec K, -, s, \fvec k}$ if $\nu_{s, \fvec k} = 2$, and $u_{s, \fvec k} d^{\dagger}_{\fvec K, +, s, \fvec k} + v_{s, \fvec k} d^{\dagger}_{\fvec K, -, s, \fvec k}$ if $\nu_{s, \fvec k} = 1$. Therefore,
\[  \langle d^{\dagger}_{\fvec K, m, s, \fvec k} d_{\fvec K, n, s, \fvec k} \rangle = \begin{pmatrix} |u_{s, \fvec k}| & u_{s, \fvec k} v^*_{s, \fvec k} \\  u^*_{s, \fvec k} v_{s, \fvec k} & |v_{s, \fvec k}|^2   \end{pmatrix}_{nm} = (M(s, \fvec k))_{nm}  \]
where we introduce a $2\times 2$ matrix $M(s, \fvec k)$ at each $s$ and $\fvec k$ to simplify the notation. Note that the momentum $\fvec k$ in the matrix $M(s, \fvec k)$ is not restricted to the first mBZ, and $M(s, \fvec k) = M(s, \fvec k + \fvec G)$ is a periodic function of $\fvec k$. Now, we consider the total energy $E =  \langle \Psi_{GS} |  H | \Psi_{GS} \rangle$ and obtain
\begin{align}
    E = &\frac1{2A} \sum_{\fvec G \neq 0} \left( \sum_{\substack{\fvec k \in \mathrm{mBZ}\\ m, n}} \Lambda^{\fvec K}_{mn}(\fvec k, \fvec k + \fvec G)  \sum_s M_{nm}(s, \fvec k) \right) \left( \sum_{\substack{\fvec k' \in \mathrm{mBZ} \\ m', n'}} \Lambda^{\fvec K}_{m' n'}(\fvec k', \fvec k' - \fvec G)  \sum_{s'} M_{n' m'}(s', \fvec k') \right) \nonumber \\
    & + \frac1{2A} \sum_{\fvec q \neq 0} \sum_{\fvec k \in \mathrm{mBZ}} \sum_{\substack{m, n\\m',n'}}\Lambda^{\fvec K}_{m n}(\fvec k, \fvec k + \fvec q) (\delta_{n m'} - M_{n m'}(s, \fvec k + \fvec q)) \Lambda^{\fvec K}_{m' n'}(\fvec k + \fvec q, \fvec k) M_{n' m}(s, \fvec k)
\end{align}
The ground state is obtained by minimizing $E$ with respect to $u_{s, \fvec k}$, $v_{s, \fvec k}$ under the constraint $|u_{s, \fvec k}|^2 + |v_{s, \fvec k}|^2 = 1$, as well as the occupation numbers $\nu_{s, \fvec k}$ that satisfy the constraint$\sum_{s, \fvec k} \nu_{s, \fvec k} = \nu N_{uc}$.

In addition, we introduce the energies of the pseudo bands by solving the equations
\begin{align}
    & \frac1A \sum_{\fvec G \neq 0}  V(\fvec G) \left( \sum_{\substack{\fvec k' \in \mathrm{mBZ} \\ m', n'}}\Lambda^{\fvec K}_{m' n'}(\fvec k', \fvec k' - \fvec G) \sum_{s'} M_{n' m'}(s', \fvec k') \right) \sum_n \Lambda^{\fvec K}_{m n}(\fvec k, \fvec k + \fvec G) \begin{pmatrix} u_{s, \fvec k} \\ v_{s, \fvec k}  \end{pmatrix}_n + \nonumber \\
    & \frac1{2A} \sum_{\fvec q \neq 0} V(\fvec q) \sum_{n, m', n'} \Lambda^{\fvec K}_{m n}(\fvec k, \fvec k + \fvec q) (\delta_{n m'} - 2 M_{n m'}(s, \fvec k + \fvec q))  \Lambda_{m' n'}^{\fvec K}(\fvec k + \fvec q, \fvec k) \begin{pmatrix} u_{s, \fvec k} \\ v_{s, \fvec k}  \end{pmatrix}_{n'} \nonumber \\
    = & E_{\alpha}(s, \fvec k) \begin{pmatrix} u_{s, \fvec k} \\ v_{s, \fvec k}  \end{pmatrix}_m \ ,
\end{align}
where $\alpha =1$ or $2$ for two bands. It is obvious that the equation above can be written as 
\[ H_{eff}(s, \fvec k) \begin{pmatrix} u_{s, \fvec k} \\ v_{s, \fvec k}  \end{pmatrix} = E_{\alpha}(s, \fvec k) \begin{pmatrix} u_{s, \fvec k} \\ v_{s, \fvec k}  \end{pmatrix}  \ ,  \]
where $H_{eff}(s, \fvec k)$ is a $2 \times 2$ matrix. The two eigenvalues $E_1(s, \fvec k)$ and $E_2(s, \fvec k)$ plotted in Fig.~\ref{Fig:BandEvolution} for different filling $\nu$. Also, the chemical potential $\mu$ is obtained from the formula $\sum_{\alpha, s, \fvec k}\Theta(\mu - E_{\alpha}(s, \fvec k)) = \nu N_{uc}$ and the occupation number $\nu_{s, k} = \sum_{\alpha} \Theta(\mu - E_{\alpha}(s, \fvec k))$.

\section{Mass of Quasiparticles}
In this section, we calculate the quasiparticle mass $m$. For this purpose, we first introduce the filling $\nu_{s, \alpha}$ for the spin $s$ and the band $\alpha$. With the chemical potential $\mu$ and the energies of the pseudo-bands $E_{\alpha}(s, \fvec k)$ obtained in the previous section, we define $\nu_{s, \alpha} = \frac1{N_{uc}}  \sum_{\fvec k} \Theta(\mu - E_{\alpha}(s, \fvec k))$.  As a consequence, the mass $m_{s, \alpha}$ for the spin $s$ and the band $\alpha$  is $m_{s, \alpha} = \dfrac{2\pi \hbar^2}{S_{uc}} \dfrac{\rmd \nu_{s, \alpha}}{\rmd \mu}$, where $S_{uc} = \dfrac{\sqrt{3}}2 |L_m|^2$ is the area of the moire unit cell.

\begin{figure}[h]
\begin{center}
\includegraphics[width=0.6\textwidth]{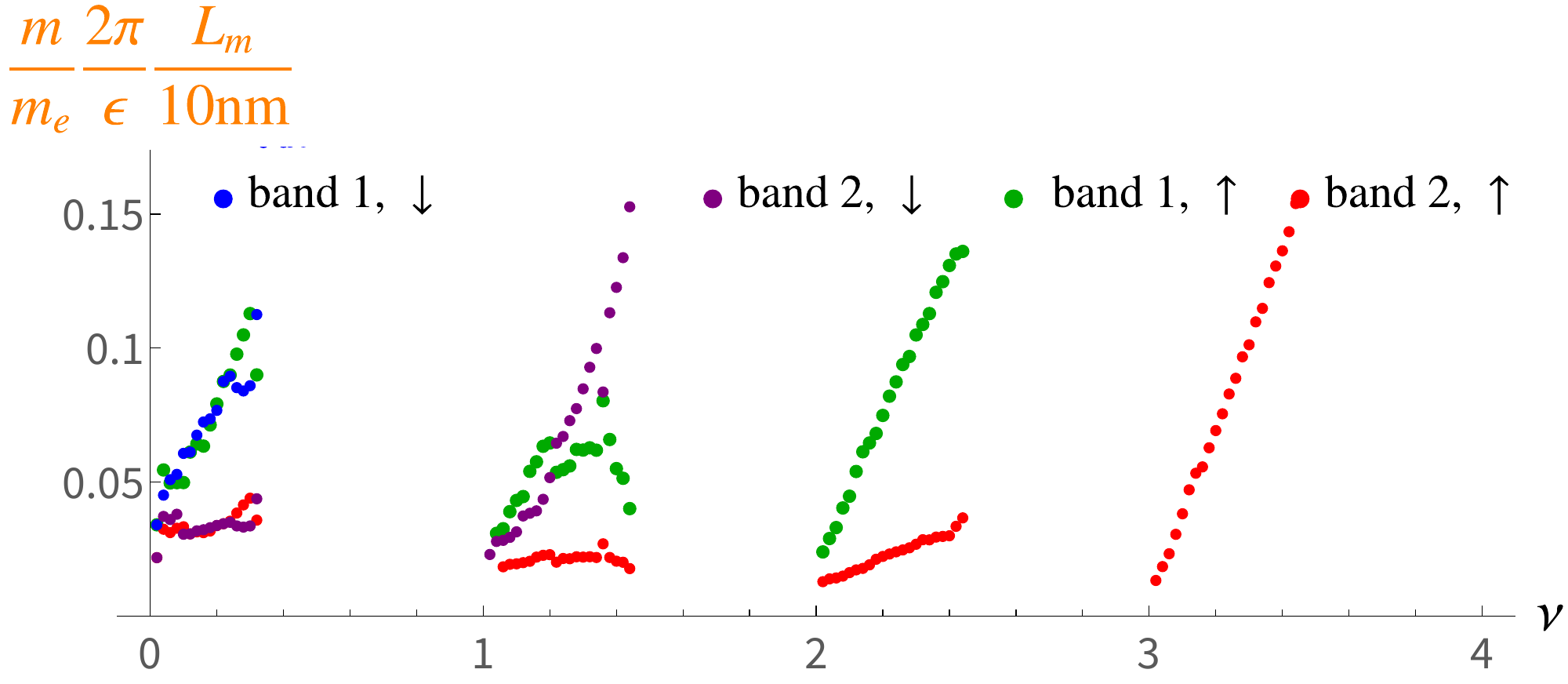} 
\includegraphics[width=0.6\textwidth]{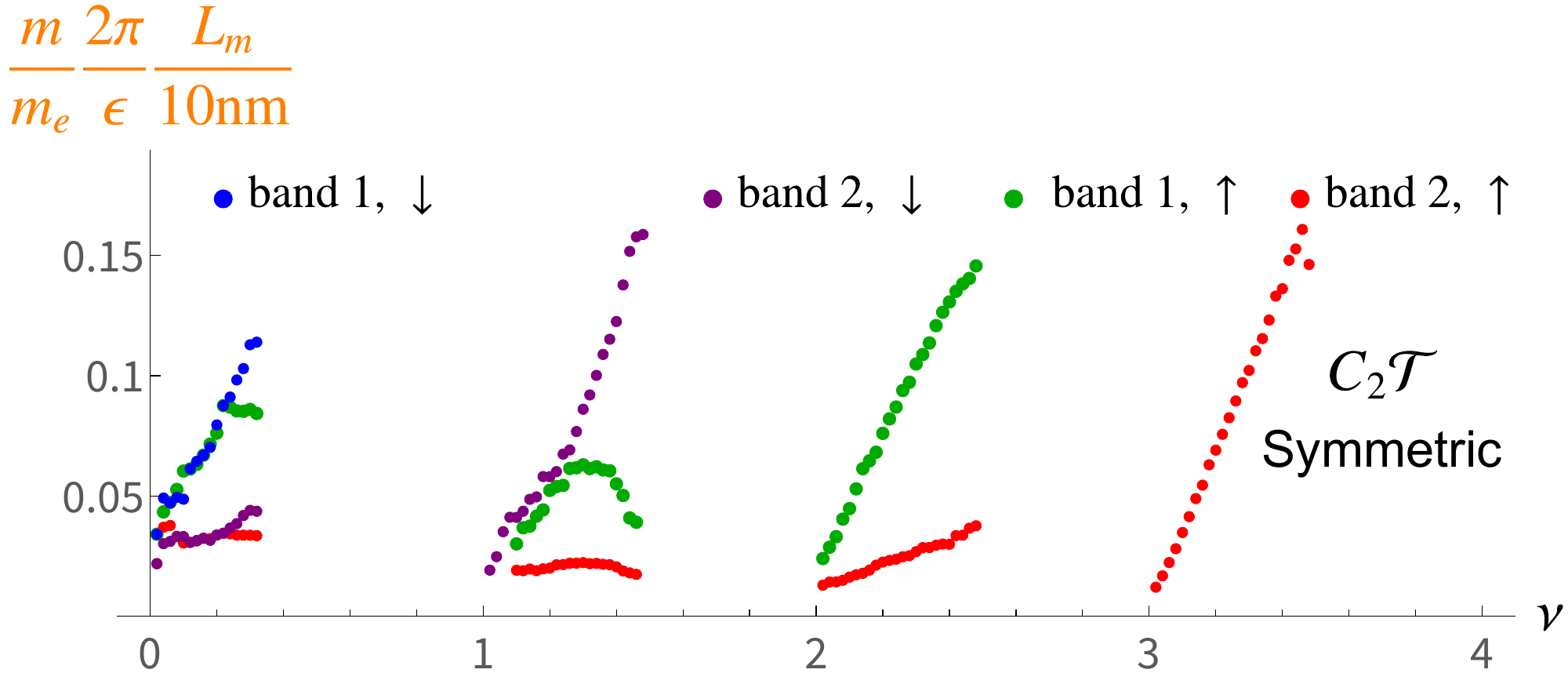} 
 \caption{The mass of the quasiparticles $m_{s, \alpha}$ for the spin $s = \uparrow$ or $\downarrow$ and the band $\alpha = 1$, or $2$ as a function  of the filling factor $\nu$ for the trial state in Eqn.~\ref{Eqn:trial} at $w_0/w_1=0.7$ when the $C_2 \mathcal{T}$ symmetry is allowed to be broken (top two panels) and when $C_2\mathcal{T}$ is enforced (bottom two panels). }  \label{FigS:Mass} 
\end{center}
\end{figure} 

Fig.~\ref{FigS:Mass} illustrates the  mass v.s.~the filling $\nu$ when $n < \nu < n + 0.4$ for the non-negative integer $n$. As discussed in the main text, when $n + 0.4 < \nu < n + 1$, the ground state and the excitations obtained from the trial state in Eqn.~\ref{Eqn:trial} may be unstable upon the residual interactions. 

\section{Symmetry Analysis of the Charge-1 Excitation Band Structure at the $\fvec \Gamma$ point}

Numerically, we observe that the minimum of the electron band structure of the states at positive filling is at the $\fvec \Gamma$- point. With particle-hole and $C_2\mathcal{T}$ symmetries, we observe a double-degeneracy at this point, and a splitting of the bands away from this point. However, the splitting does not take the Rashba form, i.e. the energy minimum remains at $\fvec \Gamma$ and not at a small but finite momentum away from $\fvec \Gamma$ as Rashba couplings would suggest. The apparent absence of a linear in $k$ term has an important consequence on the Landau-Level spectrum, which is then dominated by the $k^2$ dispersion and not by the zero-mode of the Rashba dispersion. 

In this Appendix we prove, by symmetry, that the particle-hole and $C_{2}\mathcal{T}$ symmetries stabilize Kramers-like degeneracies at the $\Gamma$ and $M$-points. We further prove that the $C_{3}$ symmetry then forbids the existence of a linear Rashba splitting away from the $\fvec \Gamma$ point, but allows the presence of a cubic splitting, giving rise to a $3$-vortex at $\fvec \Gamma$. The absence of the linear term renders the dispersion and Landau Level structure dominated by the the quadratic term. A $C_2''=C_{2x}$ rotation by $\pi$ around the $x$-axis symmetry then establishes a relation between the coefficients of the cubic terms. 

We note that the charged excitations above the insulating states satisfy a different set of symmetries than the TBG single-particle bands. In particular, the latter do not exhibit double degenearcies at the $\fvec \Gamma$ and $\fvec M$ points. In ~\cite{VK2020,AndreiV}, the expression for the charge 1 excitation above the insulating ground-state at filling $\nu$ is given as the diagonalization of the $2\times 2$ matrix:

\begin{eqnarray}
&&\mathcal{E}^{\tau}_{n' n,\pm}(\fvec k) = \frac1{2 A}   \left( \sum_{\fvec q} V(q) \sum_m \Lambda^{\tau}_{mn}(\fvec k - \fvec q, \fvec k) \Lambda^{\tau}_{n' m}(\fvec k, \fvec k - \fvec q)   \pm \frac{\nu}2 \sum_{\fvec G} V(G) \bar{\rho}_{\fvec G} \Lambda^{\tau}_{n'n}(\fvec k + \fvec G, \fvec k ) \right) , \label{Eqn:ExEneSM}
\end{eqnarray}
where $\fvec k$ is in the first mBZ while the summation over $q$ is over all possible momenta. The form factors $\Lambda^{\tau}_{n' m}(\fvec k, \fvec k - \fvec q)$ 
are given by their expression in terms of the TBG eigenstates:
\beq
\Lambda^{\tau}_{n n'}(\fvec k, \fvec k - \fvec q)  =\sum_{\fvec Q}\sum_{\alpha=1,2} u^\star_{\fvec{Q}, n,\alpha, \tau}(\fvec k) u_{\fvec{Q}, n',\alpha, \tau}(\fvec k -\fvec q)
\eneq
where $u_{ \fvec{Q}, m, \alpha, \tau} (\fvec k)$ is the eigenstate of the active band $m=\pm$ (i.e. {\em not} the Chern basis used in the main text), sublattice $\alpha = A,B\equiv 1,2$, in valley $\tau=\pm$ at momentum $\fvec k$ in the mBZ and plane-wave index $\fvec Q$ of the Bistritzer MacDonald model. The dispersion relation is the same in either valley \cite{AndreiV} hence we suppress the valley index in $\mathcal{E}^{\tau}_{n' n,\pm}(\fvec k)$. We can easily show that $\mathcal{E}_{n' n,\pm}(\fvec k)$ is Hermitian, and serves as a single particle Hamiltonian for the charge-1 excitation. The eigenstates satisfy several important symmetries which then become symmetries of the form factors. Using the gauge-fixing of \cite{AndreiIII} we have:  
\begin{eqnarray}
& C_{2}\mathcal{T}: \;\; u_{ \fvec{Q}, m, \alpha, \tau} (\fvec k)= \sigma^x_{\alpha \beta} u^\star_{ \fvec{Q}, m,\beta, \tau}(\fvec k) \implies \Lambda^{\tau} _{n n'}(\fvec k, \fvec k - \fvec q)  = \Lambda^{\tau\star}_{n n'}(\fvec k, \fvec k - \fvec q)   \nonumber \\ &
 \mathcal{P} : \;\; u_{ -\fvec{Q}, -m,\alpha, \tau} (-\fvec{k}) = - m \tau \zeta_{\fvec{Q}} u_{\fvec{Q}, m,\alpha, \tau} (\fvec k) \implies \Lambda^{\tau} _{n n'}(\fvec k, \fvec k - \fvec q)  = n n' \Lambda^{\tau}_{-n -n'}(-\fvec k, -\fvec k + \fvec q)
\end{eqnarray} 
where $\zeta_{\fvec{Q}}$ is $\pm 1$ depending on whether the plane-wave $\fvec{Q}$ is associated to the top or lower layer. We expand the charge $\pm 1$ single-particle Hamiltonian $\mathcal{E}_{n' n,\pm}(\fvec k)$ in Pauli matrices 
\beq
\mathcal{E}(\fvec k)= d_0({\fvec{k}})\sigma_0 + \sum_{i=1}^3 d_i({\fvec{k}})\sigma_i
\eneq
Using these symmetries for the BM eigenstates, we find the following properties of the charge 1 single-particle Hamiltonian $\mathcal{E}_{n' n,\pm}(\fvec k)$:
\begin{eqnarray}
& C_2 \mathcal{T}:\;\;\mathcal{E}_{n' n,\pm}(\fvec k) = \mathcal{E}^\star_{n' n,\pm}(\fvec k) \implies d_y({\fvec{k}})= 0,\;\;\forall {\fvec{k}}\nonumber \\ &  \mathcal{P} :\;\; \mathcal{E}_{n' n,\pm}(\fvec k) = n'n \mathcal{E}_{-n'- n,\pm}(-\fvec k) \implies d_0({\fvec{k}})=d_0(-{\fvec{k}}),\;\;d_{x,z}({\fvec{k}})=-d_{x,z}(-{\fvec{k}})
\end{eqnarray} 
We hence see that for $\fvec k = - \fvec k \mod \fvec G$  we find double degeneracies since $d_{x,z}$ vanish. The Kramers degeneracy is enforced by the operator product of $C_{2} \mathcal{T}$ and particle-hole $\mathcal{P}$: $\mathcal{P} C_{2} \mathcal{T}$, with the property $(\mathcal{P} C_2 \mathcal{T})^2 = -1$, due to $\mathcal{P}^2 = -1$. $C_{2} \mathcal{T}$ and $\mathcal{P}$ would then allow for a linear $\fvec k$ term away from the $\fvec \Gamma$ point. 

We now impose $C_{3}$ symmetry on the Hamiltonian. One complication is that the sewing matrix for $C_{3}$, $B^{C_{3}}({\fvec{k}})$ cannot be chosen independently of ${\fvec{k}}$ over the entire mBZ as the BM wavefunctions are topological \cite{Bernevig1}. The properties of the $C_3$ sewing matrix are~\cite{Bernevig1}
\begin{eqnarray}
& B_{mn}^{C_3}({\fvec k})u_{C_3 \fvec{Q}, n, \alpha, \tau}(C_3 \fvec{k})  = e^{i \frac{2\pi}{3} \tau \sigma^z_{\alpha \beta}} u_{\fvec{Q}, m, \beta, \tau}( \fvec{k}) \nonumber \\ &B^{C_{3}\dagger}({\fvec{k}})B^{C_{3}}({\fvec{k}}) = B^{C_{3}}({\fvec{k}})B^{C_{3}\dagger}({\fvec{k}})=B^{C_{3}}(C_{3}^2{\fvec{k}}) B^{C_{3}}({C_{3}\fvec{k}})B^{C_{3}}({\fvec{k}})= 1, \nonumber \\ & m B^{C_{3}}_{-m r} ({\fvec{k}})=-r B^{C_{3}}_{m,-r}(-{\fvec{k}})
\end{eqnarray} 
where the last property is due to the commutation with the $\mathcal{P}$ operator. Implementing these properties, the $C_{3}$ sewing matrix can be parametrized as:
\beq
B^{C_{3}}({\fvec{k}}) =\zeta_0({\fvec{k}})+ i \sigma_y \zeta_y ({\fvec{k}}),\;\;\; \zeta_0({\fvec{k}})= \zeta_0(-{\fvec{k}}),\;\;\; \zeta_y({\fvec{k}})= \zeta_y(-{\fvec{k}});\;\;\;B^{C_{3}}({\fvec \Gamma})= I
\eneq 
The last equation originates from the fact that at the $\fvec \Gamma$ point the BM eigenstates have the same eigenvalue $1$ under $C_{3}$ symmetry \cite{Bernevig1}. The charge 1 single-particle Hamiltonian $\mathcal{E}_{mn}({\fvec{k}})$ satisfies
\beq \mathcal{E}({\fvec{k}}) =B^{C_{3}\dagger}({\fvec{k}}) \mathcal{E}(C_{3}{\fvec{k}}) B^{C_{3}}({\fvec{k}})
\eneq 
which can be expanded around the $\fvec \Gamma$ point $\mathcal{E}({\fvec \Gamma+ \delta\fvec{k}}) =B^{C_{3}\dagger}({\fvec \Gamma+ \delta \fvec{k}}) \mathcal{E}(C_{3}(\fvec \Gamma + \delta \fvec{k})) B^{C_{3}}({\fvec \Gamma+ \delta \fvec{k}})$. Implementing  the symmetries we obtain:
\begin{eqnarray}
&\mathcal{E}({\fvec \Gamma+ \delta\fvec{k}}) = (d_0(\fvec \Gamma) + m (\delta k_x^2+ \delta k_y^2))\sigma_0+ (a_1 f_1(\delta {\bf k})+ a_2 f_2(\delta {\bf k})) \sigma_x + (b_1 f_1(\delta {\bf k})+ b_2 f_2(\delta {\bf k})) \sigma_z \nonumber \\& f_1(\delta {\bf k})= \delta {\bf k}_y(3\delta {\bf k}_x^2- \delta {\bf k}_y^2),\;\;\; f_2(\delta {\bf k})= \delta {\bf k}_x(3\delta {\bf k}_y^2- \delta {\bf k}_x^2)
\end{eqnarray}
where $d_0(\fvec \Gamma), m, a_1, a_2, b_1, b_2$ are constants not determined by symmetry. At $\fvec \Gamma$ we hence find a vortex of order $3$, and the band splitting is cubic. Further addition of $C_2''$ symmetry (upon gauge-fixing) gives $a_2=b_1=0$, giving the expression in polar coordinates ${\delta \fvec{k}} = \delta k (\cos\theta, \sin\theta)$:
\beq
\mathcal{E}({\fvec \Gamma+ \delta\fvec{k}}) = (d_0(\fvec \Gamma) + m + \delta k^2)\sigma_0 + a_1 \delta k^3 \cos(3\theta) \sigma_x + b_2 \delta k^3 \sin(3\theta) \sigma_z
\eneq

\section{$\fvec k \cdot \fvec p$ analysis of the direct (Hartree) term}
	
A further understanding of the quasielectron and quasihole energies can be obtained by using  $\fvec{k} \cdot \fvec p$ expansions of the BM Hamiltonian. In this way, we can obtain analytic expressions for the form factors $\Lambda(\fvec k + \fvec G, \fvec k)$ of the dispersion. While the entire form factors $\Lambda(\fvec k + \fvec q + \fvec G, \fvec k)$ and hence the entire excitation dispersion in TBG can in principle be obtained by using the $\fvec k \cdot \fvec p$-type expansions of \cite{AndreiI}, we here concentrate on the $\fvec q = 0$ term $\Lambda(\fvec{k} + \fvec G, \fvec k)$ and leave the full analytic dispersion for a future publication.   

\subsection{General Properties of the Dispersion Equation}

\subsubsection{Chiral Limit}
We first concentrate on the chiral limit;  the generalization away from the chiral limit is tedious but straightforward in the present formalism. In the $C_{2} \mathcal{T}$ and chiral symmetry $C$ gauge fixing of \cite{AndreiIII} we re-write explicitly 

\beq
\mathcal{E}({\fvec{k}})= \frac{1}{2 A} \sum_{\fvec{G}}  (\sum_{\fvec{q} \in \mathrm{mBZ}}  V({\fvec{G}+\fvec{q}})\mathrm{Tr}[\Lambda(\fvec k - \fvec q - \fvec G, \fvec k) \Lambda^\dagger(\fvec k - \fvec q - \fvec G, \fvec k)   ] \pm \nu V(G) (\sum_{k_1\in \mathrm{mBZ}}\mathrm{Tr}[\Lambda(\fvec{k}_1 - \fvec{G}, \fvec{k}_1)] \mathrm{Tr}[ \Lambda(\fvec k + \fvec G, \fvec k)])
\eneq where $\fvec{k}\in \mathrm{mBZ}$, $\nu\ge 0$ is the filling away from charge neutrality, and $\pm$ stands for electrons and holes, respectively. Negative fillings $\nu\le 0$ can be treated by many-body particle-hole conjugation. We have dropped the $\tau$ valley indices as the dispersion is identical irrespective of the valleys, and we have explicitly separated the summation over all $\fvec{q}$ in a summation over the first mBZ and over subsequent mBZs determined by $\fvec{G}$.

In the chiral limit, and in the $C_{2}\mathcal{T}, \mathcal{C}$ gauges \cite{AndreiIII}  this matrix is diagonal and independent of the valley index $\tau$. Crucially, the expression is also obtained in the periodic gauge, for which  $u_{\fvec{Q}, m, \tau}(\fvec{k}+\fvec{G}) = u_{ \fvec{Q}-\fvec{G}, m, \tau} (\fvec{k})$. We notice the following facts

\begin{itemize}
\item  Due to the decay of the eigenstates of the lowest bands with the momentum away from the $\fvec \Gamma$ point \cite{AndreiI}, only $|\fvec G|=0, \sqrt{3}$ vectors need be taken into account. There are $6$ $\fvec G$ ($|\fvec{G}|=\sqrt{3}$) vectors $\fvec{G}=\pm \fvec{G}_{1,2,3}$. Higher $\fvec {G}$ values give negligible contributions. 

\item  Due to $C_3$,  $C_2''$ symmetries, the term $\sum_{\fvec k_1 \in \mathrm{mBZ}} \Lambda(\fvec{k}_1 - \fvec{G}, \fvec{k}_1)$ only depends on $|\fvec G|$. 

\item The chiral limit dispersion is a sum of two $k$ functions. By taking the difference of electrons and holes dispersion at the same filling, or by taking the difference of the hole (or electron) dispersions at different fillings, one can obtain $\Lambda(\fvec{k} - \fvec{G}, \fvec{k})$. 
\end{itemize}

\begin{figure}[ttt]
\begin{center}
\includegraphics[width=0.98\textwidth]{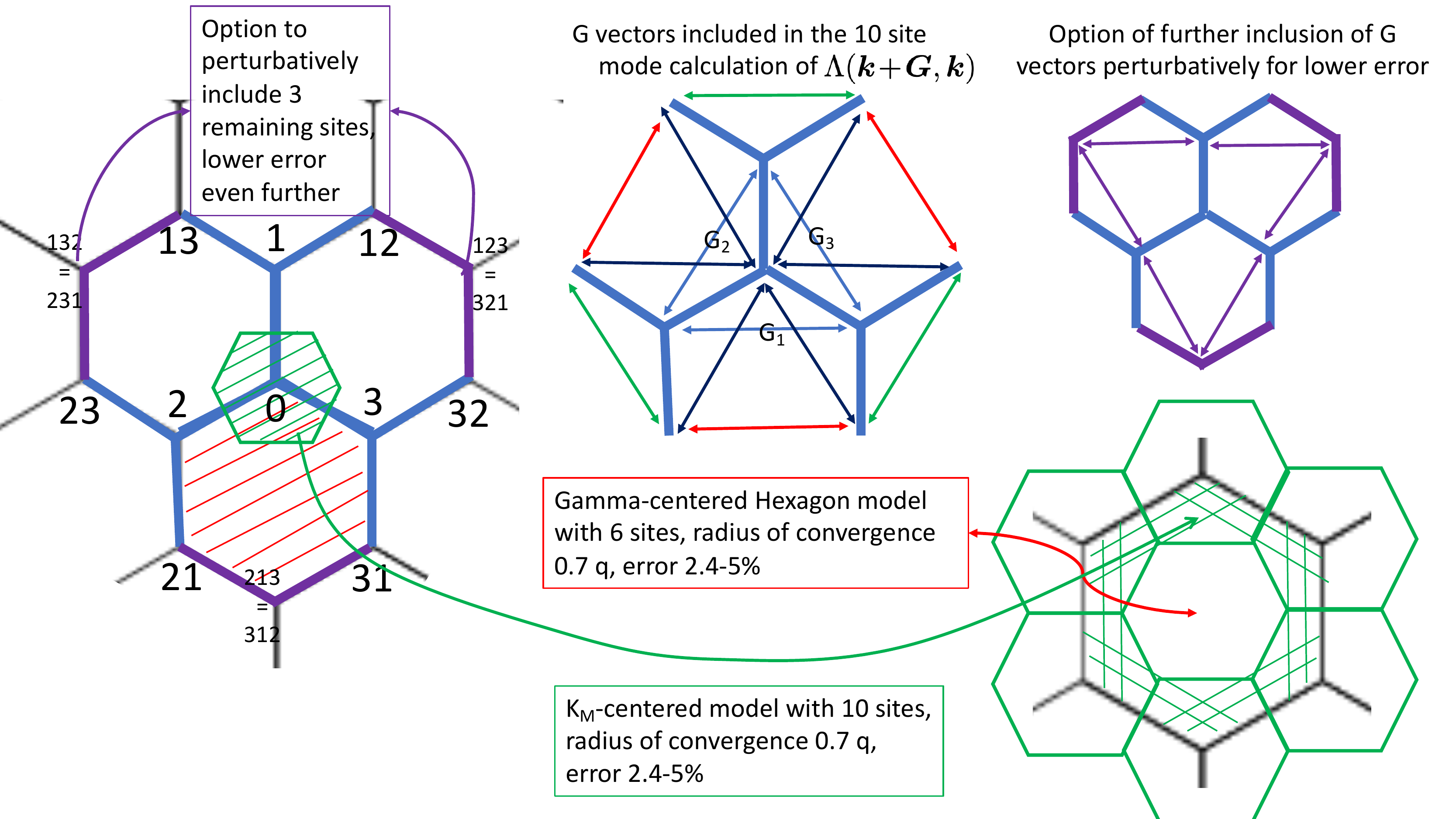} 
 \caption{ Ten site model used to compute the form factor $\Lambda(\fvec k + \fvec G, \fvec k)$ around the $\fvec{K}_M$ Dirac points produces a remarkably large radius of convergence, which includes the $\fvec{M}_M$ point, with an error of 2.5- 5\%. The form factor around the $\fvec \Gamma_M$ point will be computed using the hexagon model of \cite{AndreiI}. The option to include remaining sites $124, 132, 213$ perturbatively is also available.  \label{Lattice} } 
\end{center}
\end{figure}

\subsection{$\fvec{K}_M$ -centered Ten-Site Model}

For the physics around the $\fvec{K}_M$ point, we  first consider the tripod model which contains only the sites $0,1,2,3$ in Fig[\ref{Lattice}]. From this model, we find $\mathrm{Tr}[\Lambda(\fvec k + \fvec G, \fvec k)]$ for $|\fvec G| = \sqrt{3}$ (contributed by  the overlaps of the wavefunctions $\langle i |j\rangle$ on sites $i\ne j =1,2,3$ to be roughly $1/2$ of the numerical value,  in agreement with the order-of magnitude estimates in~\cite{AndreiI} and showing the need to include more sites. The next shell, containing the $6$ momentum site $|ij \rangle$, $i\ne j$ can also be solved analytically, and produces a much better approximation for the Hartree term $\mathrm{Tr}[ \Lambda(\fvec k + \fvec G, \fvec k)]$ for $|\fvec G | = \sqrt{3}$.  The Hamiltonian, in the basis of sites in Fig.~\ref{Lattice} ordered as
\beq
(|0   \rangle,| 1  \rangle,|2   \rangle,|   3 \rangle,|  12 \rangle,| 13  \rangle,| 23  \rangle,|  21 \rangle,| 31  \rangle,|  32 \rangle)
\eneq where each is a two-component spinor, reads

\beq
H_{\text{10 site}}= \left(\begin{matrix}
h_0 & T_1  & T_2 & T_3 &0 &0  &0 &0  &0 &0  \\
T_1& h_1 &0 & 0 & T_2&T_3  &0 &  0&0 &0  \\
T_2& 0 & h_2 & 0 &0 &0  &T_3& T_1 &0 & 0 \\
T_3& 0 &0 &h_3  &0 &0  &0 &0  &T_1 &T_2  \\
0& T_2 &0 &0   &h_{12} &  0& 0& 0 & 0& 0 \\
0& T_3 &0 & 0 &0 &  h_{13} &0 &0  & 0& 0 \\
0&0  &T_3 & 0 &0 &0  &h_{23} &0  &0 & 0 \\
0& 0 &T_1 &0  &0 & 0 & 0& h_{21} &0 &0  \\
0& 0 &0&T_1   &0 & 0 & 0 & 0 & h_{31} &0  \\
0& 0 &0 & T_2  &0 &0  &0 &0  &0 & h_{32}  
\end{matrix}\right)
\eneq where $h_0 = \fvec k \cdot \fvec \sigma; \;\; h_{i}=(\fvec k - \fvec q_i) \cdot \fvec \sigma;\;\; h_{ij} = (\fvec k- (\fvec q_i - \fvec q_j)) \cdot \fvec \sigma$ and $k$ is measured from the $\fvec{K}_M$ point. Due to the graph form of the Hamiltonian, the 10-site model is  easy to solve, giving the eigenstates: 
\beq
|i,j \rangle= (E- h_{ij})^{-1}T_j |i\rangle,\;\;\; | i\rangle =(E- h_i - \sum_{j\ne i} T_j (E- h_{ij})^{-1} T_j )^{-1} T_i |0 \rangle
\eneq 
We make two approximations: (1) keep only up to linear term in $E$, as we are interested in the  flat band energies (2) keep only small momentum $\fvec k$ away from the $\fvec{K}_M$ point.  We obtain the (Dirac) equation for $|0\rangle$
\begin{eqnarray}
&v_F \fvec k \cdot \fvec \sigma |0\rangle = E|0\rangle,\;\;\; v_F= \frac{(1-w_0^2)^2+w_0^4+w_1^4+ 4 w_0^2w_1^2 - 3 w_1^2}{(1-w_0^2)^2+3(w_0^2+ w_1^2) + 2( w_0^2+w_1^2+ 4 w_0^2w_1^2)}\nonumber \\ 
& |i,j \rangle= h_{ij}^{-1}T_j |i\rangle,\;\;\; | i\rangle =(- h_i +\sum_{j\ne i} T_j h_{ij}^{-1} T_j )^{-1} T_i |0 \rangle,\;\;\; v_F \fvec k \cdot \fvec \sigma |0 \rangle = E|0 \rangle \label{recursioneigenstates}
\end{eqnarray}  
We have two eigenstates at low energy $|{0^\pm}\rangle$ with energy $\pm v_F k $.
We gauge fix $C_{2} \mathcal{T}$ and $\mathcal{C}$ symmetries \cite{AndreiIII} by choosing $|0\rangle + (\sigma_x |0\rangle)^\star$.  We can  impose the $C_{2} \mathcal{T}$ gauge on $|0 \rangle$, and then use the new gauge fixed $|0\rangle$ in the Eq[\ref{recursioneigenstates}] to obtain the full gauge fixing.  In the chiral limit, we also have to fix the chiral gauge. This is fixed by picking an energy eigenstate $|0\rangle$ and then making the energy eigenstate of opposite energy equal $\sigma_z |0\rangle$. The Hartree form factor is
\begin{eqnarray}
&\sum_{\fvec G, |\fvec G| = \sqrt{3}} \mathrm{Tr}[ \Lambda(\fvec k + \fvec G, \fvec k)] =\frac{1}{N}\left(   \sum_{i \ne j}(  \langle i | j \rangle  + \langle ij | 0\rangle+  \langle 0 | ij \rangle  )  + \sum_{i,j,l; j \ne i, l\ne i, j \ne l}    (\langle ij |il \rangle +  \langle ij | lj \rangle)  \right) \nonumber \\ & N = \langle{0}| 0\rangle +   \sum_i \langle{i}|i \rangle   +\sum_{i\ne j}   \langle{ij}|ij\rangle
\end{eqnarray}  The terms correspond to, in succession, the light blue arrows, the dark blue arrows, the green arrows and the red arrows in upper center Fig[\ref{Lattice}] $\bf{G}$ vectors. Using the analytic eigenstates the expression for $\mathrm{Tr}[ \Lambda(\fvec k + \fvec G, \fvec k)]$ could be solved analytically.

\subsection{$\fvec{K}_M$ -Centered 16 and 19 Site Model}

One can expand further in the plane wave basis of the BM model. The 16 site model corresponds to all sites connected by the blue and yellow hoppings in   Fig[\ref{NineteenSiteModel}]. A further expansion  introduces $3$ more sites (the purple sites in the Fig[\ref{Lattice}] and is the first number of plane-waves that cannot be solved exactly, since the tree structure of the graph is broken.  The form factor $\mathrm{Tr}[\Lambda(\fvec k + \fvec G, \fvec k)]$ picks up contributions that correspond to the light blue arrows, the dark blue arrows, the green arrows and the red arrows, the yellow arrows and the black arrows in upper center Fig[\ref{NineteenSiteModel}] $G$ vectors. While the expressions for the overlaps are too large to be reproduced here, they can be obtained analytically.

\begin{figure}[ttt]
\begin{center}
\includegraphics[width=0.98\textwidth]{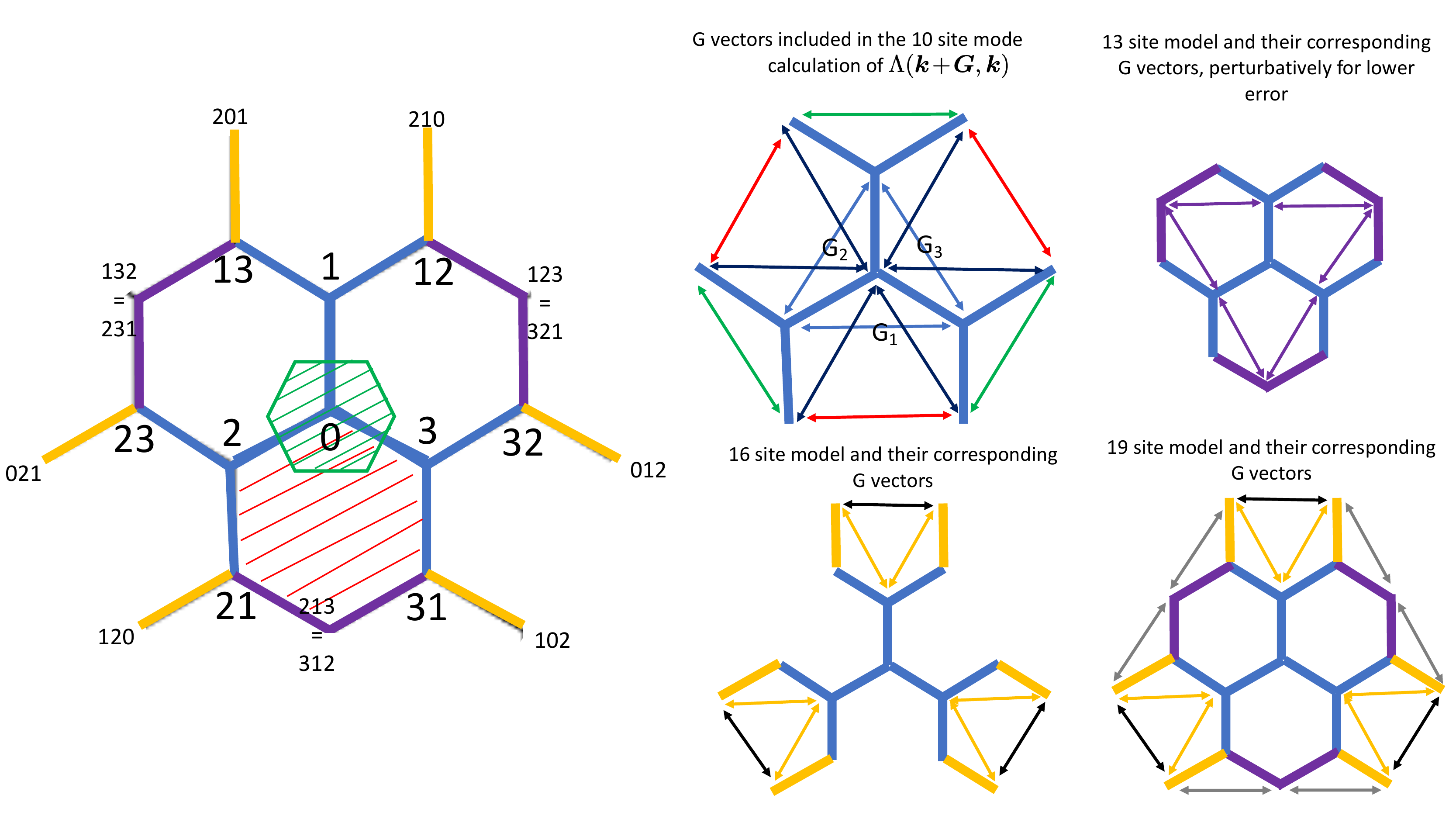} 
 \caption{Nineteen site model used to compute (approximately) the form factor $\Lambda(\fvec k + \fvec G, \fvec k)$ around the $\fvec{K}_M$ Dirac points produces a remarkably large radius of convergence, which includes the $\fvec{M}_M$ point, with an error of 0.5- 2\%. The form factor around the $\fvec{\Gamma}_M$ point will be computed using the hexagon model of \cite{AndreiI}.  The option to include further tree models is available.  \label{NineteenSiteModel} } 
\end{center}
\end{figure}

\subsection{$\fvec k\cdot \fvec p$ expansions}
The 10, 16 and 19 site models give expressions for the form factor $\mathrm{Tr}[\Lambda(\fvec{K}_M + \fvec k + \fvec G, \fvec{K}_M + \fvec k)]$ which can be analytically obtained. We present these expressions for the 10-site model, while for the 16 and 19 site models we only present the comparison with the exact numerical expression, due to the length of the analytic expressions.

\subsubsection{$\fvec{K}_M$ -centered 10-site model, Chiral limit expansion}
 The 10-site model above gives a great fit to the numerical $\mathrm{Tr}[ \Lambda(\fvec k + \fvec G, \fvec k)]$  around the $\fvec{K}_M$ point. At the $\fvec{K}_M$ point, the error is less than 2.45\%.  The form factor $\mathrm{Tr}[ \Lambda(\fvec{K}_M+ \fvec k + \fvec G, \fvec{K}_M+ \fvec k)]$  then has the expression, up to 5th order in $\fvec{k}=k(\cos \theta, \sin \theta)$ with $\fvec{k} $ measured from the $\fvec{K}_M$ point:

\begin{eqnarray}\label{MkGTensite}
& \sum_{\fvec{G}, G = \sqrt{3}} \mathrm{Tr}[ \Lambda(\fvec{K}_M+ \fvec k + \fvec G, \fvec{K}_M+ \fvec k)]  = A(w_1)+ B(w_1) k^2 + C(w_1)k^3 \sin(3 \theta) +D(w_1) k^4 + E(w_1) k^5 \sin(3\theta)\nonumber\\ &
 A(w_1)=\frac{4 w_1^2 \left(w_1^2+6\right)}{\left(w_1^2+1\right) \left(2 w_1^2+1\right)} \nonumber \\ & B(w_1)= -\frac{2 w_1^2 \left(6 w_1^{10}-9 w_1^8-164 w_1^6+357 w_1^4+393 w_1^2+27\right)}{9 \left(w_1^2+1\right)^2 \left(2 w_1^2+1\right)^2} \nonumber \\ & C(w_1) =-\frac{4 w_1^2 \left(12 w_1^{10}-161 w_1^8+396 w_1^6+153 w_1^4-333 w_1^2+333\right) }{27 \left(w_1^2+1\right)^2 \left(2 w_1^2+1\right)^2}\nonumber \\ & D(w_1) = \frac{2 w_1^2 \left(12 w_1^{18}-168 w_1^{16}+951 w_1^{14}-1071 w_1^{12}-9805 w_1^{10}+7137 w_1^8+18333 w_1^6+3033 w_1^4-2763 w_1^2-243\right)}{81 \left(w_1^2+1\right)^3 \left(2 w_1^2+1\right)^3} \nonumber \\ & E(w_1) = \frac{20 w_1^2 \left(42 w_1^{18}-553 w_1^{16}+2085 w_1^{14}-7888 w_1^{12}+21855 w_1^{10}+14853 w_1^8-61668 w_1^6-20466 w_1^4+17361 w_1^2+729\right)}{243 \left(w_1^2+1\right)^3 \left(2 w_1^2+1\right)^3}\end{eqnarray} 
 The radius of convergence is large, more than $0.3$. I find the radius of convergence to be of about $0.7 q_1$, with 6\% error on the $\fvec \Gamma - \fvec K$ line, smaller on the $\fvec K - \fvec M$ line. 
 
 This expression points to two generic features of the Hartree term. First, we notice that the coefficient of $k^2$ is negative, implying that the maximum is at the $\fvec{K}_M$ point.  Second, along the $\fvec{K}_M -\fvec{M}_M$ line $\theta = - \pi/6$ the $-C(w_1)$ coefficient of the $k^3$ term is positive. Along the $\fvec{K}_M - \fvec \Gamma_M$  line $\theta= -\pi/2$ the coefficient $C_{w_1}$ of the $k^3$ term is negative. Hence the dispersion on $\fvec{K}_M- \fvec \Gamma_M$ is downward steeper than that on the $\fvec{K}_M - \fvec{M}_M$, another feature observed in the numerical plots. The plot of the 10-site $\fvec k \cdot \fvec p$ expansion can be seen in Fig.~\ref{MkGPlots} upper left. It is a rather good approximation (2.5\%) around the $\fvec K$ point, and indeed around the $\fvec M$ point (5\%) in absolute value, but fails to describe the second derivative (mass) around the $\fvec M$ point. To solve this, we look at the expansions of the 16 and 19 site model. 

\begin{figure}[ttt]
\begin{center}
\includegraphics[width=0.98\textwidth]{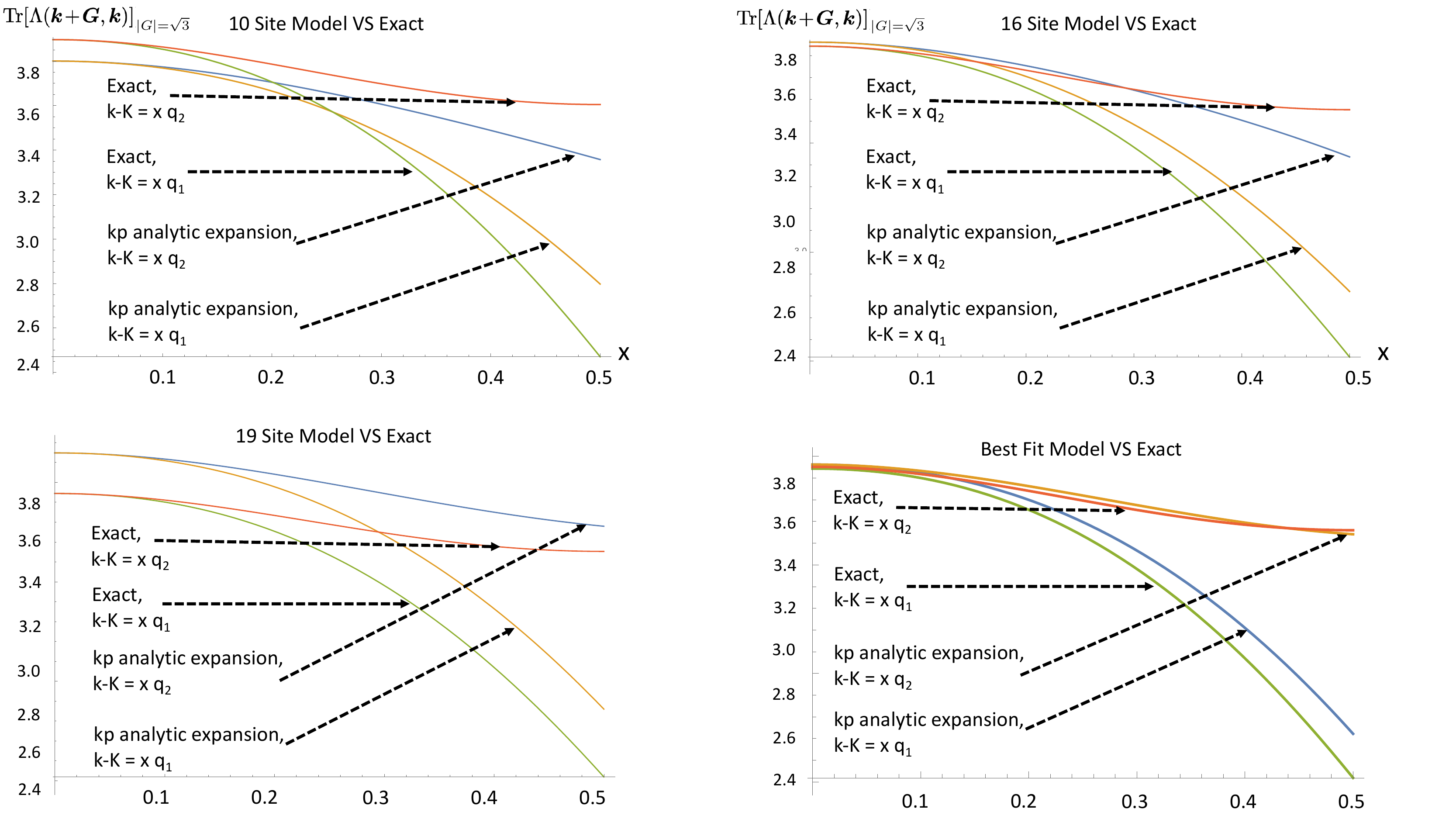} 
 \caption{ The exact  numerical $\mathrm{Tr}[\Lambda(\fvec{K}_M + \fvec{k} + \fvec G, \fvec{K}_M + \fvec{k})]$ and the approximate $k\cdot p$ approximations of the several models. Upper left: $10$-site $k\cdot p$ model  of Eq[\ref{MkGTensite}]. While providing a good approximation (within $2.5\%$ of the exact value) for both the value and the second derivative at the $\fvec K$ point and for $|\fvec k|\le 0.1$, it fails close to the $\fvec M$ point: while the absolute value at the $\fvec M$ point of Eq[\ref{MkGTensite}] is within $6$ percent of the exact value, the second derivative has the wrong sign. \label{MkGPlots} } 
\end{center}
\end{figure}

\subsubsection{$\fvec{K}_M$ -centered 16 Site model, Chiral Limit Expansion} \label{16sitemodelexpansion}

The 16-site model, whose analytic expressions are too long to write down, gives a good fit to the numerical $\mathrm{Tr}[ \Lambda(\fvec{K}_M+ \fvec k + \fvec G, \fvec{K}_M+ \fvec k)]$ around the $\fvec{K}_M$ point. At the $\fvec{K}_M$ point, the error is less than 0.45\%.  The form factor $\mathrm{Tr}[\Lambda(\fvec{K}_M+ \fvec k + \fvec G, \fvec{K}_M+ \fvec k)]$  expression, up to $5$th order in $\fvec{k}$ measured from the $\fvec{K}_M$ point can be obtained analytically but is too long to be reproduced here. We  plot  the $\fvec{k}\cdot \fvec{p}$ dispersion in Fig\ref{MkGPlots}] upper right. It is a  fantastic approximation ($<$0.5\%) around the $\fvec K$ point, and indeed around the $\fvec M$ point (3\%) in absolute value, but again fails to describe the second derivative (mass) around the $\fvec M$ point. To solve this, we need to add the shells denoted in purple in Fig[\ref{Lattice} ]

\subsubsection{$\fvec{K}_M$ -centered 19 Site model, Chiral Limit Expansion}\label{19sitemodelexpansion}
The 19-site model, whose analytic expressions are too large to provide, gives a good  fit to the numerical $\mathrm{Tr}[ \Lambda(\fvec{M}_M+ \fvec k + \fvec G, \fvec{M}_M+ \fvec k)]$ around the $\fvec M$ point. At the $\fvec{K}_M$ point, the error is  however, larger than that of the 15 site model, due to the approximations used in obtaining the analytic form of the model, which is no longer a tree.  The form factor $\mathrm{Tr}[ \Lambda(\fvec{K}_M+ \fvec k + \fvec G, \fvec{K}_M+ \fvec k)]$ can be analytically obtained up to $5$th order in $\fvec{k}$  measured from the $\fvec{K}_M$ point. The plot of the  $\fvec{k}\cdot \fvec{p}$ dispersion can be seen in Fig[\ref{MkGPlots}] lower left. It gives the correct mass ($<$0.5\%) around the $M$ point but is shifted from the exact numerical plot vertically, due to the error induced by treating the 3 new added sites perturbatively.

\subsubsection{Best Fit Model}

We now try the "best fit mode", which corresponds to taking the $k^0$ and $k^2$ terms from the $16$ site exact model, and the $k^{3,4,5}$ from the $19$ site model.  The plot can be seen in Fig\ref{MkGPlots}] lower left. It gives the correct mass (<0.5\%) around the $M$ point  and the correct  (less than $0.5\%$ error) value at the $\fvec{K}_M$ point. 

\subsubsection{Fitting to a nearest neighbor model}

We now ask if there exist \emph{an energy fit} to a \emph{triangular} lattice nearest neighbor model:

\beq \label{NNfit1}
 \sum_{\fvec{G}, |\fvec G| = \sqrt{3}} Tr[ \Lambda( \fvec k + \fvec G, \fvec k)] = 3c_1 +2 c_2 \sum_{j=1,2,3} \cos(\fvec k \cdot \fvec a_j), \;\;\; a_1= \frac{4\pi}{3 k_\theta} (0,1),\;\; a_2=  \frac{4\pi}{3 k_\theta}(\frac{\sqrt{3}}{2}, \frac{1}{2}),\;\; a_3=  \frac{4\pi}{3 k_\theta}(-\frac{\sqrt{3}}{2}, \frac{1}{2})
\eneq Since we have the expansion of $ \Lambda( \fvec k + \fvec G, \fvec k)$ around the $\fvec{K}= \frac{1}{3} (\fvec{G}_2+ \fvec{G}_3)$ point, we can check the fitting with the above dispersion. In particular, performing an expansion of the above we have
\beq
 \sum_{\fvec{G}, |\fvec G| = \sqrt{3}} Tr[ \Lambda(\fvec{K}_M+ \fvec k + \fvec G, \fvec{K}_M+ \fvec k)] = 3 c_1-3 c_2+\frac{4}{3} \pi ^2 c_2 k^2  +\frac{8 \pi ^3 c_2 k^3 \sin (3 \theta)}{9 \sqrt{3}}-\frac{4}{27} \pi ^4 c_2 k^4 -\frac{8 \pi ^5 c_2 k^5 \sin (3 \theta)}{81 \sqrt{3}}
\eneq for $\vec{k} = k(\cos\theta, \sin\theta)$. We now note that:
\begin{itemize}

\item  In a NN expansion, the sign of the $k^2 $ and $k^3$ terms are identical. The sign of the $k^4$ and $k^5$ term are also identical, and opposite to that of the $k^2$ and $k^3$

\item In a NN expansion, the ratio of the $k^3, k^4, k^5$ coefficients to the $k^2$ coefficient is universal $\left\{\frac{2 \pi  }{3 \sqrt{3}},-\frac{\pi ^2}{9},-\frac{2 \pi ^3 }{27 \sqrt{3}}\right\} \approx \{1.2092, -1.09662, -1.32604\}$

\item Using Eq[\ref{MkGTensite}] of the 10-site model, these ratios are  $\{0.846515, 0.0100919, -0.344459\}$; Hence they differ from the NN triangular model.

\item Using the 19 site approximate model, we obtain for these ratios $\{ 1.14121,  -0.203833, -0.352002\}$ which is much closer to the triangular NN model, including the value of the $k^3$ term and the sign of the $k^4$ and $k^5$ terms (if not their values) 

\item Using the 16 site exact model, we obtain for these ratios $\{0.957131, 0.044555, -0.83897\}$  which again is relatively to the triangular NN model, 

\item Using the 12 site approximate model, we obtain for these ratios $\{1.05389, -0.115228, 0.164263 \}$ which again has the  $k^3/k^2$ ratio within 15\% of the NN triangular model. 

\item Using the best fit model(16 site model around $\fvec K$, 19 site model around $\fvec M$), we obtain for these ratios $\{1.20719, -0.215618, -0.372354\}$. This is in perfect agreement with the NN approximation for the ratio of the $k^3/k^2$ term, with error less than 0.2\%. The sign of the $k^4$ and $k^5$ terms is correct. By fitting our analytic form for $\Lambda(\fvec k + \fvec G, \fvec k)$ to the dispersion Eq[\ref{NNfit1}] we obtain $c_2= -0.23445 c_1$.

\end{itemize}

\subsection{Away from the Chiral Limit}

Away from the chiral Limit, the two degenerate bands will split. We can obtain the non-chiral limit form of the $\Lambda( \fvec k + \fvec G, \fvec k)$ matrix easily, at least for the 10-site model. The form factor matrix is, with  $ \Lambda( \fvec k + \fvec G, \fvec k) = v_{D_1} k \sigma_x + O(k^2)$, where the Dirac velocity around the $\fvec K$ point can be obtained as
\beq
v_{D_1}=\frac{2 w_0 w_1 \left(4 w_0^2+11 w_1^2\right)}{(w_0-1) (w_0+1) \left(8 w_0^2 w_1^2+3 w_0^4+w_0^2+2 w_1^4+3 w_1^2+1\right)}
\eneq 
The dispersion around the $\fvec K$ point becomes "Rashba"-like away from the chiral limit with the velocity given above.

\end{widetext}

\end{document}